\pdfoutput=1

\documentclass[11pt]{article}

\usepackage[final]{acl}

\usepackage{times}
\usepackage{latexsym}
\usepackage{multirow}
\usepackage{graphicx}
\usepackage{amsmath}
\usepackage{amssymb}
\usepackage{booktabs}
\usepackage{subcaption}
\usepackage{caption}
\usepackage{bbding}
\usepackage{colortbl}
\usepackage{tcolorbox}
\usepackage[T1]{fontenc}
\usepackage{xcolor}         
\usepackage{color}
\usepackage{colortbl,array}
\usepackage[rgb]{xcolor}
\usepackage{tikz}
\usetikzlibrary{tikzmark}
\usepackage{arydshln}
\makeatletter

\usepackage[normalem]{ulem}
\useunder{\uline}{\ul}{}
\newcommand*\myfontsize{%
  \@setfontsize\myfontsize{8}{8}%
}

\makeatother
\newcommand{\mytextbox}[2]{\tikzmarknode[draw=#1,thick,inner sep=2pt]{test}{\myfontsize #2}}
\definecolor{myred}{rgb}{0.7, 0.3, 0.0}
\definecolor{myblue}{RGB}{235,245,250}
\definecolor{mygreen}{HTML}{056b34}
\definecolor{myorange}{HTML}{ff8800}
\definecolor{mypurple}{HTML}{8400ff}
\definecolor{mypink}{HTML}{f7acb9}

\newcommand{\green}[1]{\mytextbox{mygreen}{\textbf{\textcolor{mygreen}{#1}}}}

\newcommand{\purple}[1]{\mytextbox{mypurple}{\textbf{\textcolor{mypurple}{#1}}}}

\newcommand{\ie}{\textit{i.e.}}
\newcommand{\eg}{\textit{e.g.}}
\usepackage{algorithm}
\usepackage{algorithmic}

\usepackage[T1]{fontenc}

\usepackage[utf8]{inputenc}

\usepackage{microtype}
\usepackage{bbold}

\usepackage{inconsolata}

\title{ReasonRank: Empowering Passage Ranking with Strong Reasoning Ability}

\author{Wenhan Liu$^1$, Xinyu Ma$^2$, Weiwei Sun$^3$, Yutao Zhu$^1$, \textbf{Yuchen Li}$^2$ \\ 
\textbf{Dawei Yin}$^2$ \and \textbf{Zhicheng Dou}$^{1}$\thanks{Corresponding author.} \\
$^1$Gaoling School of Artificial Intelligence, Renmin University of China \\
$^2$Baidu Inc., Beijing, China \;\;
$^3$Carnegie Mellon University\\
\texttt{lwh@ruc.edu.cn, xinyuma2016@gmail.com, dou@ruc.edu.cn}
}

\begin{document}
\maketitle

\begin{abstract}
Large Language Model (LLM) based listwise ranking has shown superior performance in many passage ranking tasks. 
With the development of Large Reasoning Models (LRMs), many studies have demonstrated that step-by-step reasoning during test-time helps improve listwise ranking performance. 
However, due to the scarcity of reasoning-intensive training data, existing rerankers perform poorly in many complex ranking scenarios, and the ranking ability of reasoning-intensive rerankers remains largely underdeveloped.
In this paper, we first propose an automated reasoning-intensive training data synthesis framework, which sources training queries and passages from diverse domains and applies DeepSeek-R1 to generate high-quality training labels. 
To empower the listwise reranker with strong reasoning ability, we further propose a two-stage training approach, which includes a cold-start supervised fine-tuning (SFT) stage and a reinforcement learning (RL) stage.
During the RL stage, we design a novel multi-view ranking reward tailored to the multi-turn nature of listwise ranking.
Extensive experiments demonstrate that our trained reasoning-intensive reranker \textbf{ReasonRank} outperforms existing baselines significantly and also achieves much lower latency than the pointwise reranker.
Our codes are available at \url{https://github.com/8421BCD/ReasonRank}.

\end{abstract}

\section{Introduction} \label{sec:intro}

Passage ranking plays a crucial role in Information Retrieval (IR) by reranking initial retrieved passages to improve the quality of search results.
Recently, Large Language Models (LLMs) have demonstrated impressive capabilities in zero-shot passage ranking~\cite{rankgpt}.
Among these LLM-based ranking approaches, listwise ranking has emerged as particularly effective, as it evaluates and ranks a list of passages simultaneously for a given query, capturing global relevance patterns more comprehensively than pointwise ranking~\cite{rankt5} and pairwise ranking~\cite{qin2023large}. It has demonstrated state-of-the-art (SOTA) performance on numerous IR benchmarks~\cite{rankgpt}.

Recently, test-time reasoning~\cite{r1} has been shown to substantially improve LLM performance across a wide range of challenging NLP tasks.
By explicitly generating step-by-step reasoning during inference, LLMs can produce more accurate answers.
Such reasoning ability is also desirable for passage reranking, where understanding query intent and reasoning across multiple passages are critical to accurate ranking.
Motivated by this, recent studies~\cite{rank1, rank-k} have attempted to inject reasoning capabilities into passage rerankers, resulting in improved performance compared to directly predicting final ranking results.


\begin{figure}[t]
	\centering
	\includegraphics[width=1\linewidth]{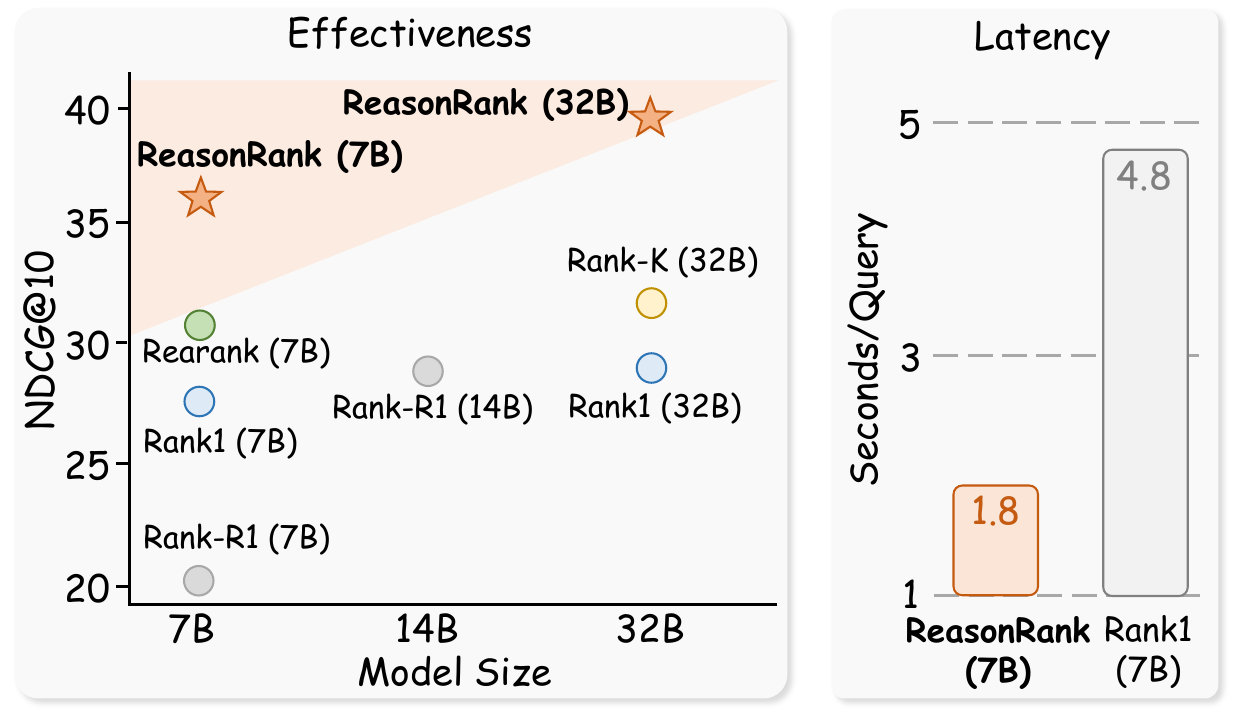}
	\caption{The left part shows the average NDCG@10 on the BRIGHT benchmark by reranking the top 100 passages retrieved by ReasonIR (8B). The right part compares the ranking latency of Rank1 (7B) and our ReasonRank (7B)  on the Earth Science dataset.}\label{fig:intro}        
\end{figure}

However, due to \textit{scarcity of reasoning-intensive training data}, existing rerankers are primarily trained on traditional web search datasets such as MSMARCO~\cite{MSMARCO}, where relevance is often determined by shallow lexical or semantic matching. However, many real-world search platforms (\eg, StackExchange) typically involve complex reasoning-intensive queries whose relevant passages are expected to provide key evidence or solutions with similar methodologies. As a result, rerankers trained on traditional web search datasets often fail to generalize to these complex search scenarios. Although collecting new training data via human annotation could alleviate this issue, it is prohibitively expensive and difficult to scale. These challenges create bottlenecks in training rerankers for complex search scenarios.

To alleviate this issue, we propose an automated data synthesis framework to construct high-quality reason-intensive training data without any human involvement. 
Specifically, we first collect diverse user queries from StackExchange platforms and well-established public datasets, covering four representative scenarios: complex QA, coding, math, and web search.
Then we apply the strong reasoning model DeepSeek-R1 to automatically select positive passages and hard negative passages to construct training passages, and generate reasoning chains and gold ranking lists as the training labels. 
To ensure the quality of R1-generated training labels, we design a self-consistency data filtering mechanism that keeps only gold rankings where R1-selected positive passages are consistently ranked at top positions.
Finally, we obtain 13K high-quality and diverse training data.

\begin{figure}[t]
	\centering
	\includegraphics[width=1\linewidth]{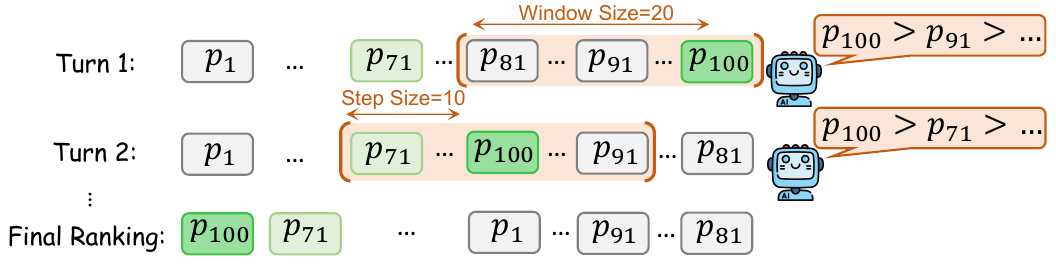}
	\caption{Ranking 100 passages based on sliding window strategy. Green squares represent relevant passages.}\label{fig:intro_sliding}  
    \vspace{-2mm}
\end{figure}

Based on our training data, we further propose a two-stage training framework to train our reasoning-intensive reranker ReasonRank. 
In the first stage, we apply cold-start SFT to teach the backbone LLM to reason over a list of passages and generate a reranked list.
In the second stage, we apply RL to further optimize reranker's listwise ranking ability using task-specific rewards.
Existing RL-based rerankers~\cite{coranking,rearank} typically use ranking metrics (\eg, NDCG) as rewards to evaluate reranked list.
However, due to the limited context length of LLMs, existing listwise ranking approaches~\cite{rankgpt,rankzephyr} often adopt a sliding-window strategy which ranks a subset of passages iteratively to obtain final global ranking (see Figure~\ref{fig:intro_sliding}). 
Under such strategy, higher NDCG on local ranking window does not guarantee that more relevant passages are recalled into later windows and therefore may fail to yield a high-NDCG global ranking (details will be discussed in Section~\ref{sec:multi_view_rl}).
To address this limitation, we propose a multi-view ranking reward that considers local ranking metric while promoting the recall of relevant passages, thereby improving final ranking metric.

Extensive experiments on reasoning-intensive IR benchmarks BRIGHT~\cite{bright} and R2MED~\cite{r2med} demonstrate the SOTA performance of ReasonRank. 
Notably, as shown in Figure~\ref{fig:intro}, the 7B-scale and 32B-scale of our ReasonRank outperform the previous state-of-the-art model by 3 and 5 points on BRIGHT, respectively, and our ReasonRank performs more efficiently than the pointwise reranker Rank1~\cite{rank1}.

The contributions of the paper are:

(1) To address the data scarcity problem in training reasoning-intensive ranking models, we design an automated data synthesis framework and generate 13K high-quality and diverse reasoning-intensive training data.

(2) We propose a two-stage training framework, which includes a cold-start SFT strategy for reasoning pattern learning and a multi-view ranking reward-based RL approach for further ranking ability enhancement.

(3) Extensive experiments on the reasoning-intensive IR benchmarks BRIGHT and R2MED demonstrate the effectiveness and efficiency advantages of our ReasonRank.

\section{Related Work}
\paragraph{LLMs for Ranking}
The application of Large Language Models (LLMs) to ranking tasks has revolutionized traditional IR~\cite{llm4ir_survey,inters,rankllm}. Current LLM-based ranking methods can be categorized into three paradigms: pointwise, pairwise, and listwise methods. Pointwise methods~\cite{liang2022holistic, sachan2022improving, demorank, fan2024rrnorm, HuangZDW25, liu2024personalize} evaluate each query-document pair independently. While computationally efficient, such a method lacks cross-document comparison. Pairwise methods~\cite{qin2023large} compare document pairs to establish relative relevance. Such a method usually suffers from efficiency issues due to the large number of pairwise comparisons. Listwise methods~\cite{rankgpt, listt5, fullrank, acurank, tourrank, pe_rank, fan2025llm} leverage LLM to rerank a passage list. Through global passage comparison, listwise ranking has achieved SOTA performance on many IR benchmarks. 

\paragraph{Reasoning for IR}
Recently, test-time reasoning has been shown as a strong ability to improve LLMs and also applied to reasoning-intensive IR scenarios~\cite{bright, reasonir, tongsearch, tfrank}, particularly in passage ranking. For example, Rank1~\cite{rank1} and Rank-K~\cite{rank-k} propose to distill the reasoning chain of Deepseek-R1 into rerankers. \citet{rank-r1} proposes using the RL algorithm GRPO to optimize a setwise reranker. While effective, they only use traditional web search datasets (\eg, MSMARCO) for training, which makes rerankers perform poorly in reasoning-intensive benchmarks. In this paper, we propose to synthesize reasoning-intensive training data from diverse domains and propose a two-stage training approach to enhance ranking performance.


\begin{figure*}[!tb]
  \centering
  \vspace{-3mm}
  \includegraphics[width=.9\linewidth]{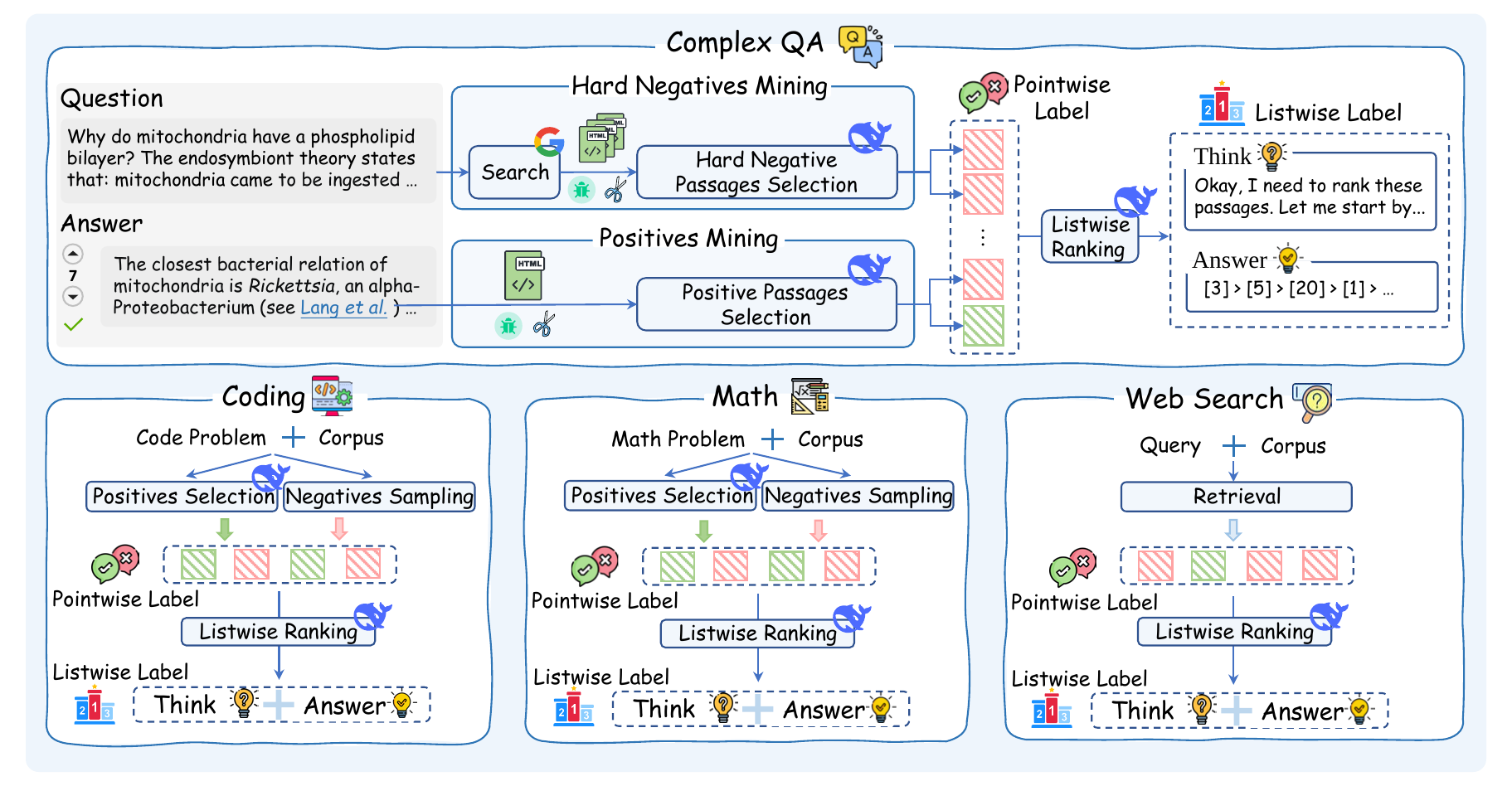}
  \caption{An overview of reasoning-intensive ranking data synthesis on four domains.}\label{fig:model1}
  \vspace{-2mm}
\end{figure*}

\section{Preliminaries}

\paragraph{Listwise Ranking}
Passage ranking aims to rerank a list of retrieved passages $[p_1, \ldots, p_N]$ based on their relevance to a query $q$. The listwise ranking approach takes both the query $q$ and a passage list as input and outputs a reranked sequence of passage IDs (\eg, [3] > [1] > \dots). Through comparing multiple passages, listwise ranking demonstrates SOTA performance in IR benchmarks~\cite{rankgpt}. Due to the limited context length of many LLMs, listwise ranking usually applies a sliding window strategy (see Figure~\ref{fig:intro_sliding}) to process a subset of passages iteratively. This strategy uses a window size $w$ and step size $s$ to promote relevant passage from back to the front. Following existing studies~\cite{rankgpt,rankzephyr}, we set $N = 100$, $w = 20$, and $s = 10$ in this paper.

\paragraph{Reasoning-based Listwise Ranking}
Reasoning-based listwise reranker takes the query and a passage list as input, and produces a structured output that includes both reasoning traces (enclosed by tags \texttt{<think>} and \texttt{</think>}) and the reranked list such as ``[3] > [1] > \dots'' (enclosed by tags \texttt{<answer>} and \texttt{</answer>}).

\section{Methodology}
To train a reasoning-intensive reranker, we propose an automated data synthesis framework to collect high-quality reasoning-intensive training data and design a two-stage training framework for empowering the ranking ability.


\subsection{Reasoning-intensive Data Synthesis}
To mitigate the scarcity issue of reasoning-intensive training data, we introduce an automated framework for synthesizing training data. 
Existing reasoning-intensive IR benchmarks~\cite{bright, r2med, rar-b} mainly evaluate queries from three main domains: \textbf{complex QA}, \textbf{math}, and \textbf{coding}. 
Following these benchmarks, we also use these three kinds of queries, as well as \textbf{web search} query (to ensure ReasonRank's performance on short search queries), to synthesize our training data. The overall process is shown in Figure~\ref{fig:model1}.
We further design a self-consistency data filtering mechanism to remove low-quality training data. Next, we will detail each part.

\subsubsection{Complex QA}
Complex QA queries are typically long and involve complex reasoning. We source complex QA training queries from StackExchange\footnote{\url{https://archive.org/download/stackexchange}}, which contains expert-level questions that require a deep understanding and complex reasoning. We select six sub-domains: Biology, Earth Science, Economics, Robotics, StackOverflow, and Sustainability.

\textbf{Positive Passages Selection.} Training data for listwise ranking usually consists of passage lists with relevant (positive) and irrelevant (negative) passages. Relevant passages for a complex QA query often contain key concepts that help answer the query. For StackExchange questions, relevant evidence often appears in external documents linked in accepted answers. To mine the positives of each query, we first obtain the gold answer that is accepted by the user and has at least one URL, crawl the linked document of each URL, and split them into passages. We then employ a strong model, Deepseek-R1 (R1), to perform listwise \emph{positive passages selection}, which takes the query, gold answer, and a candidate passage list as input and outputs the IDs of positive passages (prompt in Figure~\ref{fig:positive_selection}). This yields the positive set $P^+$, with the remaining passages treated as negatives $P^-$. Note that providing a passage list enables R1 to perform comparative relevance assessment, while the gold answer helps understand the query and judge the relevance.

\textbf{Hard Negative Passages Selection.} 
To further enrich the passage list, we mine hard negatives—passages that are topically related to the query yet provide no value for answering it. Specifically, we retrieve top-10 documents using the Google Search API, split them into passages, and apply R1 to perform listwise \emph{hard negative passages selection} (prompt in Figure~\ref{fig:hardneg_selection}), producing a hard negative set $P_{\text{hard}}^-$.


\textbf{Listwise Ranking.} 
Finally, we combine all the collected passages $P^+$, $P^-$, and $P_{\text{hard}}^-$ into a passage list capped at 20 passages (\ie, the sliding window size) and apply R1 for \emph{listwise ranking}, generating both reasoning chains and reranked lists (prompt in Figure~\ref{fig:r1_listwise}). So far, we have obtained two types of training labels: (1) pointwise binary labels (1 for $P^+$ and 0 for $P^-$ and $P_{\text{hard}}^-$) and (2) listwise labels with reasoning chains and gold rankings. Both training labels will be used in our two-stage training approach. We do not provide the gold answer when generating reasoning chains with R1, to avoid reasoning chains' reliance on the gold answer and ensure consistency with ReasonRank training (same for the following coding and math domains).


\subsubsection{Coding}
For coding-based queries, relevant passages typically share the same algorithmic strategy or similar logic. We use coding problems from the LeetCode dataset\footnote{\url{https://huggingface.co/datasets/greengerong/leetcode}} as training queries, and adopt the corpus of the LeetCode dataset from the BRIGHT benchmark to mine positives and negatives.

Mining positives for coding-type queries is non-trivial, as relevant passages may not exhibit high lexical similarity to the query. To mine positives, we first retrieve a candidate set of $K=40$ passages using a strong dense retriever E5-Mistral-7B-Instruct~\cite{e5-mistral}, and then apply R1 for positive passage selection (prompt in Figure~\ref{fig:positive_selection_coding}). We sample negatives from the remaining passages and combine them with the selected positives to form a passage list of 20 passages. Finally, we apply R1 for listwise ranking to obtain the listwise label, including a reasoning chain and a gold ranking list.

\subsubsection{Math}
Math queries often require retrieving problems with similar solutions or related theorems. Following BRIGHT, we consider two math passage ranking tasks with different relevance definitions: (1) \textbf{Math-Problem}, which ranks problem–solution pairs based on similarity in problem-solving logic to the query, and (2) \textbf{Math-Theorem}, which ranks mathematical theorems according to whether they are applicable for solving the query. We use math problems from the MATH dataset~\cite{math} as training queries. For the Math-Problem task, we adopt the STEM problem–solution corpus constructed by~\cite{bright}; for the Math-Theorem task, we use 20K theorem statements from ProofWiki\footnote{\url{https://proofwiki.org}} as the passage corpus.

Following the same procedure as in coding queries, we retrieve the top-$K$ passages using E5-Mistral-7B-Instruct, apply R1 for positive passage selection (prompts in Figure~\ref{fig:positive_selection_math} and~\ref{fig:positive_selection_math-theorem}), sample negatives to construct passage lists, and finally use R1 for listwise ranking to obtain listwise labels.

\subsubsection{Web Search}
Beyond the above three domains, we also incorporate web search queries to ensure the model’s ranking ability in simple search tasks. We sample 4K queries from the MSMARCO training set whose passages have been annotated with pointwise relevance labels. Following previous studies~\cite{rankvicuna, fullrank}, we use BM25 to retrieve top-20 passages and apply R1 for listwise ranking to obtain listwise labels.

\subsubsection{Self-Consistency Data Filtering}
Through the above data synthesis process, we have obtained our training queries, passage lists, and corresponding labels -- specifically, the \textbf{pointwise labels and listwise labels} generated by R1. To ensure the quality of training labels, we propose a self-consistency data filtering mechanism. Inspired by the LLM self-consistency framework proposed in~\cite{self-consistency}, we hypothesize that the labels derived from R1 with higher self-consistency should correspond to higher label quality. Thus, we compute the ranking metric NDCG@10 for the gold ranking lists from the listwise labels, using the pointwise labels as ground truths. We then remove all training samples with NDCG@10 values below a predefined threshold $\alpha$. Through this process, we finally obtain a quality-filtered dataset for subsequent model training. Detailed information about the dataset is shown in Table~\ref{tab:statistics}.

\begin{figure*}[!tb]
  \centering
  \vspace{-2mm}
  \includegraphics[width=.9\linewidth]{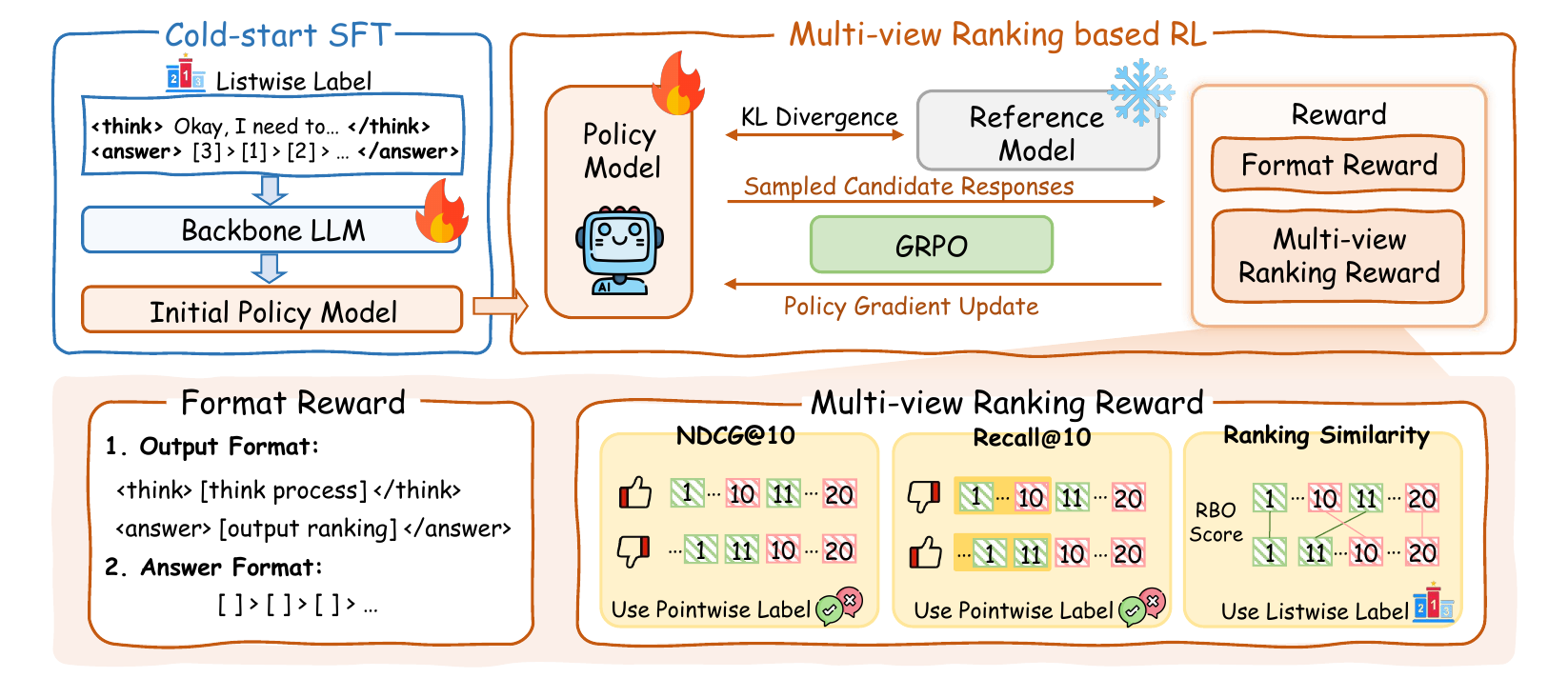}
  \caption{An overview of our two-stage training framework.}\label{fig:model2}  
  \vspace{-2mm}
\end{figure*}

\subsection{Two-stage Training Framework}
In this section, based on our synthesized data, we present a two-stage training approach (see Figure~\ref{fig:model2}) to empower our reasoning-intensive listwise reranker with strong ranking ability. Specifically, we first utilize the listwise labels to perform supervised fine-tuning (SFT) on the backbone LLM, endowing it with initial reasoning capability for listwise ranking tasks. Then, we design a novel multi-view ranking reward tailored to the characteristics of sliding-window-based listwise ranking, and apply reinforcement learning (RL) to further enhance the reranker’s ranking performance.

\paragraph{Cold-Start SFT}
To enable the LLM to learn reasoning for listwise ranking, we leverage the listwise labels, which contain both the reasoning chains and the reranked passage list, for SFT. The input to the backbone LLM comprises a query paired with a passage list. The training process optimizes the model by minimizing the standard language modeling loss, as formalized in the equation below:
\begin{equation}
\label{equ:loss1}
\mathcal{L} = -\sum_{i=1}^{|y|} \log(P_{\theta}(y_i \mid x, y_{<i})),
\end{equation}
where $x$ and $y$ represent the input prompt and the listwise label, respectively. The input prompt is shown in Figure~\ref{fig:reasonrank_prompt}.

\paragraph{Multi-view Ranking based RL} \label{sec:multi_view_rl}
In this section, we employ RL to further help the reranker discern subtle ranking differences and optimize for ranking metric.

The reward signal serves as the optimization objective that directly influences the policy model’s training efficacy. Previous studies~\cite{coranking, rearank} merely adopt the ranking metric NDCG@10 as the reward signal. However, as we mentioned in Section~\ref{sec:intro}, such a \emph{single-turn reward} is suboptimal for sliding window-based listwise ranking. This is because the sliding window strategy entails \emph{multi-turn sequential ranking}, where maximizing NDCG@10 for a local ranking window does not necessarily ensure the optimal NDCG@10 of the final ranking. For example, with window size of 20 and step size of 10, ranking two relevant passages (among 20 passages) at positions 1 and 11 yields a higher NDCG@10 score (0.61) than placing them at positions 9 and 10 (0.36). Nevertheless, ranking these passages at positions 9 and 10 ensures both remain within the top-10 and propagate to subsequent windows, potentially leading to better NDCG@10 in the final ranking. 

Thus, in addition to NDCG@10, we propose incorporating the metric Recall@10 as part of our ranking reward. Recall@10 is computed as the proportion of relevant passages retrieved in the top 10 positions of the current ranking window. Moreover, compared to using pointwise labels for NDCG@10 calculation, we contend that the gold list in our listwise labels contains more granular ranking signals. Consequently, we adopt the Rank-Biased Overlap (RBO) metric~\cite{rbo}, which measures listwise ranking similarity, as an additional supplementary ranking reward. The formula is shown in Equation~(\ref{eq:rbo}).

A combination of NDCG@10, Recall@10, and RBO with pre-defined parameters $\phi$ and $\gamma$ forms our \textbf{multi-view ranking reward $R^{\text{m}}$}:
\begin{equation}
R^{\text{m}} = \text{NDCG@10} + \phi \cdot \text{Recall@10} + \gamma \cdot \text{RBO}. 
\end{equation}
Furthermore, to ensure a proper output structure, we implement format rewards considering two types of formats: (1) Output format, which ensures the presence of both the \texttt{<think>} and \texttt{<answer>} tags; and (2) Answer format, which validates that the content enclosed in the \texttt{<answer>} tags adheres to the specified ranking format (\eg, [4] > [2] \dots).

The final reward $R$ is computed as follows:
\begin{equation}
\begin{aligned} 
R &=
\begin{cases}
R^{\text{m}}, & \text{Both formats are correct}, \\
0, & \text{Only answer format is incorrect}, \\
-1, & \text{Output format is incorrect}.
\end{cases} 
\end{aligned}\label{eqn:reward}
\end{equation}

%
We employ the GRPO algorithm~\cite{grpo} for optimization with the reward $R$. Details of the GRPO algorithm are shown in Appendix~\ref{app:rl}.

\section{Experiment}

\begin{table*}[!t]
\small
\centering
\vspace{-2mm}
\setlength{\tabcolsep}{1.15mm}{
\begin{tabular}{lccccccccccccc}
\toprule
\textbf{Models} & \textbf{Econ.} & \textbf{Earth.} & \textbf{Rob.} & \textbf{Bio.} & \textbf{Psy.} & \textbf{Stack.} & \textbf{Sus.} & \textbf{Leet.} & \textbf{Pony} & \textbf{AoPS} & \textbf{TheoQ.} & \textbf{TheoT.} & \textbf{Avg.} \\ \midrule
ReasonIR (8B) & 32.65 & 43.00 & 20.82 & 43.49 & 39.57 & 30.96 & 27.34 & 31.69 & 19.55 & 7.37 & 33.93 & 36.68 & 30.59 \\ \midrule
\multicolumn{14}{l}{\textit{\textbf{Non-reasoning reranker}}} \\
rankT5 (3B) & 11.35 & 22.11 & 10.94 & 13.62 & 11.40 & 11.35 & 15.96 & 27.45 & \underline{38.05} & \underline{9.24} & 18.27 & 9.46 & 16.60 \\
RankZephyr (7B) & 19.87 & 17.36 & 12.35 & 34.90 & 24.72 & 13.35 & 22.34 & \underline{29.29} & 32.37 & 6.05 & 28.98 & 30.07 & 22.64 \\ \midrule
\multicolumn{14}{l}{\textit{\textbf{Reasoning reranker}}} \\
Rank-R1 (7B) & 20.19 & 27.91 & 18.13 & 36.70 & 30.22 & 11.32 & 29.28 & 17.13 & 9.35 & 3.23 & 14.43 & 28.98 & 20.57 \\
Rank-R1 (14B) & 27.39 & 38.73 & 23.11 & 44.45 & 37.10 & 27.82 & 36.77 & 21.27 & 19.23 & 8.80 & 31.66 & 39.53 & 29.66 \\
Rank1 (7B) & 25.34 & 38.91 & 16.77 & 39.91 & 35.32 & 24.81 & 33.47 & 12.71 & 28.14 & 2.58 & 30.67 & 38.18 & 27.23 \\
Rank1 (32B) & 25.43 & 37.97 & 17.14 & 42.43 & 34.85 & 23.82 & 31.18 & 12.20 & \textbf{40.98} & 4.79 & 29.28 & 39.95 & 28.34 \\
Rearank (7B) & 30.65 & 40.27 & 27.16 & 46.94 & 40.77 & 26.12 & 36.26 & \textbf{30.55} & 22.84 & 7.31 & 32.28 & 39.86 & 31.75 \\
Rank-K (32B) & 30.09 & 39.78 & 26.60 & 50.64 & 43.45 & 29.87 & 35.17 & 27.23 & 22.80 & 7.61 & 37.06 & 41.00 & 32.61 \\ 
\rowcolor{myblue} ReasonRank (7B) & \underline{35.05} & \underline{47.75} & \underline{31.22} & \underline{56.70} & \underline{47.77} & \underline{32.47} & \underline{40.87} & 23.17 & 24.95 & 7.68 & \textbf{39.49} & \underline{41.80} & \underline{35.74} \\
\rowcolor{myblue} ReasonRank (32B) & \textbf{36.64} & \textbf{48.90} & \textbf{33.88} & \textbf{58.17} & \textbf{53.27} & \textbf{38.68} & \textbf{45.97} & 25.78 & 20.93 & \textbf{9.41} & \underline{38.46} & \textbf{46.28} & \textbf{38.03} \\ \bottomrule
\vspace{-2mm}
\end{tabular}}
\caption{The results (NDCG@10) on the BRIGHT benchmark. All baselines rerank the top 100 passages retrieved by ReasonIR. The top two rerankers are highlighted in \textbf{bold} and \underline{underlined}.}\label{tab:bright}
\end{table*}

\begin{table*}[!t]
\small
\centering
\setlength{\tabcolsep}{2.7mm}{
\begin{tabular}{lccccccccc}
\toprule
\textbf{Models} & \textbf{Biology} & \textbf{Bioin.} & \textbf{MedS.} & \textbf{MedE.} & \textbf{MedD.} & \textbf{PMCT.} & \textbf{PMCC.} & \textbf{IIYiC.} & \textbf{Avg.} \\ \midrule
E5-mistral (7B) & 18.28 & 41.47 & 41.01 & 6.44 & 11.38 & 19.81 & 30.97 & 21.37 & 23.84 \\ \midrule
\multicolumn{10}{l}{\textit{\textbf{Non-reasoning reranker}}} \\
RankT5 (3B) & 13.20 & 32.84 & 23.38 & 2.09 & 4.18 & 0.61 & 14.55 & 12.39 & 12.91 \\
RankZephyr (7B) & 22.86 & 43.07 & 48.24 & 6.97 & 10.45 & 26.64 & 7.78 & 14.59 & 22.58 \\ \midrule
\multicolumn{10}{l}{\textit{\textbf{Reasoning reranker}}} \\
Rank-R1 (7B) & 34.04 & 51.63 & 50.96 & 12.79 & 21.96 & 34.79 & 31.70 & 25.08 & 32.87 \\
Rank-R1 (14B) & 38.78 & 53.81 & 57.94 & 15.15 & 25.49 & 40.57 & 42.52 & \underline{29.63} & 37.99 \\
Rank1 (7B) & 32.64 & 55.57 & 54.74 & 12.78 & 19.98 & 34.40 & 30.17 & 18.15 & 32.30 \\
Rank1 (32B) & 31.84 & \underline{61.65} & 59.74 & \underline{16.56} & \underline{26.90} & \textbf{41.31} & \underline{45.56} & 29.49 & 39.13 \\
Rearank (7B) & 38.35 & 50.91 & 59.73 & 14.06 & 19.34 & 37.49 & 34.29 & 26.27 & 35.06 \\
Rank-K (32B) & 32.94 & 53.94 & 51.47 & 11.41 & 22.93 & 34.44 & 38.38 & 26.24 & 33.97 \\ 
\rowcolor{myblue}ReasonRank (7B) & \textbf{46.80} & 59.70 & \underline{60.11} & 16.48 & 24.92 & 39.21 & 39.13 & \textbf{29.85} & \underline{39.53} \\
\rowcolor{myblue}ReasonRank (32B) & \underline{45.56} & \textbf{67.73} & \textbf{63.45} & \textbf{18.90} & \textbf{30.60} & \underline{41.08} & \textbf{46.11} & 29.35 & \textbf{42.85} \\ \bottomrule
\end{tabular}}
\caption{The results (NDCG@10) on R2MED benchmark. All models rerank E5-mistral-retrieved top-100 passages.}\label{tab:r2med}
\vspace{-2mm}
\end{table*}

\subsection{Experimental Setup}

\paragraph{Evaluation Datasets}
We choose two reasoning-intensive IR benchmarks, BRIGHT~\cite{bright} and R2MED~\cite{r2med}, for evaluation. BRIGHT consists of 12 datasets from diverse domains and is widely used to test the reasoning-intensive IR models. R2MED is a benchmark designed for reasoning-driven medical retrieval, which contains 8 datasets. 

As for the BRIGHT benchmark, we use a strong reasoning dense model ReasonIR (8B)~\cite{reasonir} as the initial retriever. Following previous studies~\cite{reasonir, rank-k}, we use the GPT4-rewritten queries (provided in the test set) as test queries due to better retrieval performance and use the original test query for passage reranking. As for the R2MED benchmark, we use E5-Mistral-7B-Instruct as the initial retrievers and original test query for retrieval and reranking. We rerank the top 100 retrieved passages and use NDCG@10 as the evaluation metric.

\paragraph{Baselines}
We compared two types of rerankers: non-reasoning rerankers and reasoning rerankers. As for non-reasoning rerankers, we choose RankT5 (3B)~\cite{rankt5} and RankZephyr (7B)~\cite{rankzephyr}. As for reasoning rerankers, we compare with Rank1 (7B, 32B)~\cite{rank1}, Rank-R1 (7B, 14B)~\cite{rank-r1}, Rearank (7B)~\cite{rearank} and Rank-K (32B)~\cite{rank-k}. The baseline details are shown in Appendix~\ref{app:baselines}. 

We use Qwen2.5-7B-Instruct and Qwen2.5-32B-Instruct to train our ReasonRank. Experiments based on other backbone LLMs are provided in Section~\ref{subsec:different_backbones}. Due to space limitations, we put the whole implementation details in the Appendix~\ref{app:implementation_details}.


\subsection{Main Results}
We evaluate our ReasonRank (7B and 32B) on BRIGHT and R2MED, and show the results in Table~\ref{tab:bright} and Table~\ref{tab:r2med}, respectively. From the results, we have the following observations:

\textbf{(1) Our ReasonRank (7B and 32B) demonstrates superior performance compared with all baselines on the average of two benchmarks.} 
Notably, our ReasonRank (32B) outperforms the best baseline Rank-K (32B) on BRIGHT by about 5 points and Rank1 (32B) on R2MED by about 4 points. Besides, our ReasonRank (7B) even outperforms the 32B-scale baselines significantly. For example, it surpasses Rank-K (32B) by 3 and 9 points on BRIGHT and R2MED, respectively. These results demonstrate the effectiveness of our ranking data synthesis and two-stage training framework.

\textbf{(2) Existing baselines struggle in reasoning-intensive reranking.}
In the BRIGHT benchmark, the baselines, except for Rank-K (32B), can hardly improve the initial retrieval results. In the R2MED benchmark, the two non-reasoning rerankers underperform the retriever E5-Mistral on the Avg. metric. This suggests that traditional training data and existing training methods struggle to produce an effective reasoning-intensive reranker. 

\textbf{(3) The performance of the reranker scales with the model size.}
For example, on the BRIGHT benchmark, Rank-R1 (32B) exceeds Rank-R1 (7B) by 9 points (Avg.), and our ReasonRank (32B) exceeds ReasonRank (7B) by about 2.3 points (Avg.). This indicates that larger models have stronger reasoning and ranking capabilities.

\subsection{Ablation Study}
We conduct ablation studies to evaluate the contribution of different components from three aspects: (1) training data, (2) training approach, and (3) \textit{w/o} reasoning. Experiments are conducted on the BRIGHT benchmark using ReasonRank (7B). Overall results are reported in Table~\ref{tab:ablation}, with per-dataset results and detailed analyses provided in Table~\ref{tab:ablation_detailed} and Appendix~\ref{app:ablation}. We summarize our findings as follows.

\textbf{(1) Training Data.}
Training on only the MSMARCO subset leads to a substantial performance drop (5.66 points), highlighting the importance of diverse, reasoning-intensive training data. Removing self-consistency filtering (\textit{w/o} Self-Consistency) also degrades performance, demonstrating its effectiveness.


\textbf{(2) Training Approach.}
We ablate the two-stage training framework by removing cold-start SFT (``\textit{w/o} SFT (only RL)'') and RL (``\textit{w/o} RL (only SFT)'').
Removing cold-start SFT leads to a substantial performance drop (7.05 points), highlighting its critical role in enabling the model to acquire reasoning ability for listwise ranking.
Removing RL also degrades performance (2.59 points), indicating additional gains from reinforcement learning.
Moreover, replacing the proposed multi-view ranking reward with NDCG@10 alone (``\textit{w/o} $R^m$'') further harms performance, validating the effectiveness of our reward design.


\textbf{(3) W/o Reasoning.}
To validate the effectiveness of reasoning, we train non-reasoning rerankers that directly output rankings using only the gold rankings in our training data. We experimented with two training strategies: only SFT and SFT + RL (\ie, GRPO). From the table, both approaches (32.96 and 33.60) perform significantly lower than our ReasonRank (35.74), confirming that explicit reasoning is essential for listwise ranking.


\begin{table}[]
\small
\centering
\vspace{-0.2cm}
\begin{tabular}{p{4cm}p{1.2cm}l}
\toprule
\textbf{Models} & \multicolumn{2}{c}{\textbf{BRIGHT Average}} \\ \midrule
ReasonRank (7B) & \textbf{35.74} & \multicolumn{1}{c}{-} \\ \midrule
\multicolumn{3}{l}{\textit{\textbf{Training data}}} \\
\quad only MSMARCO & 30.08 & -5.66 \\
\quad \textit{w/o} Self-Consistency & 34.20 & -1.54 \\ \midrule
\multicolumn{3}{l}{\textit{\textbf{Training approach}}} \\
\quad \textit{w/o} SFT (only RL) & 28.69 & -7.05 \\
\quad \textit{w/o} RL (only SFT) & 33.15 & -2.59 \\
\quad \textit{w/o} $R^m$ & 34.20 & -1.54 \\ \midrule
\textit{\textbf{w/o reasoning}} & \multicolumn{1}{l}{} &  \\
\quad Only SFT & 32.96 & -2.78 \\
\quad SFT + RL & \multicolumn{1}{l}{33.60} & -2.14 \\ \bottomrule
\end{tabular}
\caption{Ablation study based on ReasonRank (7B).}\label{tab:ablation}
\end{table}

\subsection{Traditional IR Benchmark}
To evaluate the generalization of ReasonRank, we conduct experiments on the traditional IR benchmark BEIR~\cite{beir}. Considering the large size of some test sets in BEIR, following the previous study~\cite{rankgpt}, we selected 7 datasets with a smaller number of queries from BEIR. We choose BM25 as our retriever and compare it with several competitive baselines: RankZephyr (7B), Rank-R1 (14B), Rank1 (32B), and Rank-K (32B). From the results in Table~\ref{tab:beir_avg}, we can see that ReasonRank (32B) outperforms all baselines, demonstrating its strong generalization ability on the traditional IR benchmark. Besides, the gap of our ReasonRank over the baselines is smaller than that on BRIGHT and R2MED. This might be because the queries in BEIR are much simpler and do not require complex reasoning. Results for each dataset are shown in Table~\ref{tab:beir_detailed}.

\begin{table}[!t]
\small
\centering
\vspace{-0.2cm}
\begin{tabular}{p{4cm}p{1.2cm}l}
\toprule
\textbf{Models} & \multicolumn{2}{c}{\textbf{BEIR Average}} \\ \midrule
BM25 & 43.74 & \multicolumn{1}{c}{-} \\ \midrule
RankZephyr (7B) & 54.14 & +10.40 \\
Rank-R1 (14B) & 54.61 & +10.87 \\
Rank1 (32B) & 50.99 & +7.25 \\
Rearank (7B) & 54.29 & +10.54 \\
Rank-K (32B) & 48.34 & +4.60 \\ 
\rowcolor{myblue}ReasonRank (7B) & 54.35 & +10.61 \\
\rowcolor{myblue}ReasonRank (32B) & \textbf{55.44} & \textbf{+11.70} \\ \bottomrule
\end{tabular}
\caption{Averaged NDCG@10 on 7 BEIR datasets, including Covid, DBPedia, SciFact, NFCorpus, Signal, Robust04, and News.}\label{tab:beir_avg}
\end{table}


\begin{figure}[!t]
    \centering
	\includegraphics[width=0.8\linewidth]{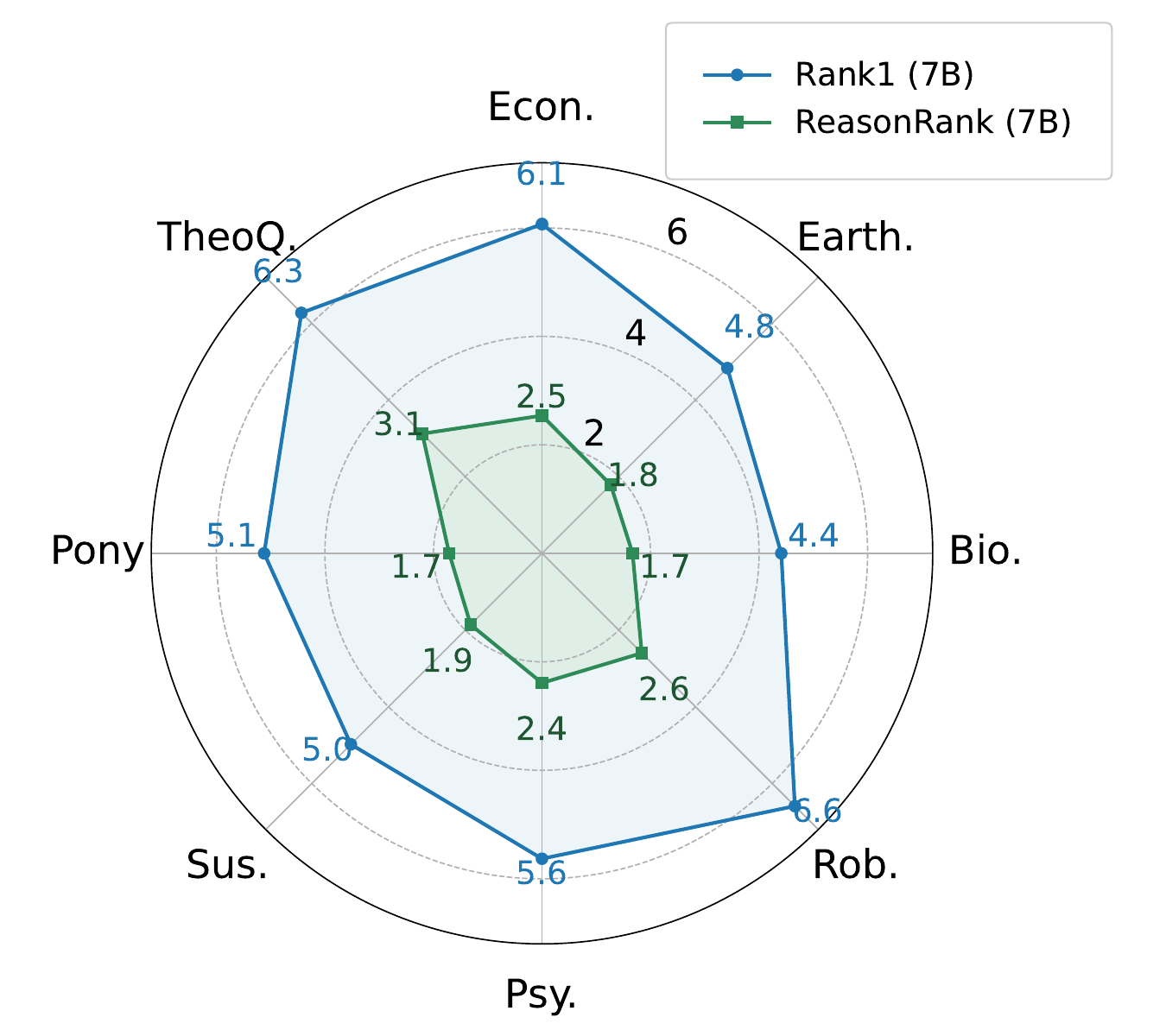}
	\caption{Ranking latency (seconds per query) of Rank1~(7B) and ReasonRank~(7B) on eights datasets.}\label{fig:latency}
\end{figure}

\subsection{Efficiency Analysis}
ReasonRank improves ranking performance through reasoning but also introduces additional latency. In this section, we test the time latency of ReasonRank (7B) and compare it with the pointwise reranker Rank1 (7B). We select 8 datasets from BRIGHT and rerank the top-100 ReasonIR-retrieved passages on 4*A800 80G GPUs. As shown in Figure~\ref{fig:latency}, surprisingly, our listwise ReasonRank is 2-2.7× faster than pointwise Rank1, which is contrary to conclusions from non-reasoning rerankers~\cite{setwise}. This efficiency stems from Rank1 generating a reasoning chain for each passage, while ReasonRank processes multiple passages at a time with only one reasoning chain, significantly decreasing the number of output tokens.

\section{Conclusion}
In this paper, we propose ReasonRank, a state-of-the-art reasoning-intensive passage reranker. To train ReasonRank, we first propose a reasoning-intensive ranking data synthesis framework and then design a two-stage training framework, which includes a cold-start SFT and a multi-view ranking-based RL. Extensive experiments demonstrate the superior performance as well as the efficiency of our ReasonRank.

\section*{Limitations} \label{limitation}
Despite the superior performance achieved by ReasonRank, we believe that our work still has some limitations: (1) ReasonRank does not include non-reasoning type data during training, which results in its inability to seamlessly switch between reasoning and non-reasoning modes when faced with search scenarios of varying difficulty. In the future, we plan to introduce non-reasoning type data into the training set to enhance ReasonRank’s flexibility in handling different scenarios (like Qwen3~\cite{qwen3}). (2) ReasonRank still relies on the sliding window strategy for passage reranking. Existing studies~\cite{fullrank} have demonstrated that LLMs have strong full-list ranking capabilities (i.e., directly ranking 100+ passages in one forward pass), which exhibit both superior efficiency and effectiveness compared to sliding window approaches. In future work, we plan to explore reasoning-intensive listwise reranking based on full ranking to further improve the scalability and performance of our framework.

\section*{Acknowledgments}
The work was supported by National Natural Science Foundation of China No. 62272467, and was partially done at the Engineering Research Center of Next-Generation Intelligent Search and Recommendation, MOE.


\clearpage
\appendix

\section{Benchmarks and Baselines}

\subsection{Benchmarks}
In our experiments, we utilize three IR benchmarks for evaluation: BRIGHT~\cite{bright}, R2MED~\cite{r2med}, and BEIR~\cite{beir}. Each of these benchmarks is instrumental in assessing different aspects of our retrieval models, ensuring a comprehensive evaluation across varied scenarios. The detailed introduction of each benchmark is as follows:

\begin{itemize}
    \item BRIGHT: The BRIGHT benchmark is designed to challenge information retrieval systems with complex queries that necessitate deep reasoning capabilities. Unlike traditional benchmarks that focus on simple keyword or semantic matches, BRIGHT comprises 1,384 queries drawn from diverse domains such as economics, psychology, mathematics, and coding. These queries are sourced from both naturally occurring data and curated human inputs, making the benchmark a robust test for advanced retrieval models. The requirement for thorough reasoning to identify relevant documents makes BRIGHT particularly demanding, as it pushes models to go beyond surface-level matching and engage in deeper cognitive processing.
    
    \item R2MED: R2MED is a specialized benchmark for medical information retrieval, emphasizing the importance of reasoning in retrieving relevant medical information. It features 876 queries that cover three main tasks: Q\&A reference retrieval, clinical evidence retrieval, and clinical case retrieval. These tasks span a wide array of medical scenarios and body systems, highlighting the necessity for retrieval systems to understand and process medical information accurately. R2MED underscores the critical role reasoning plays in medical decision-making, challenging retrieval systems to move beyond simple lexical or semantic matching to support complex clinical judgments and evidence-based practice.
    
    \item BEIR: BEIR is a comprehensive benchmark designed to evaluate the performance of information retrieval systems across 18 diverse datasets. These datasets encompass a variety of tasks, including fact-checking, question answering, and more, allowing for a thorough assessment of a model's generalization capabilities. BEIR is particularly focused on the ability of retrieval systems to adapt across different domains and tasks, emphasizing the need for a balance between performance and computational efficiency. By encouraging the development of robust and adaptable retrieval systems, BEIR serves as a pivotal benchmark in the advancement of information retrieval technologies.
\end{itemize}

\subsection{Baselines} \label{app:baselines}
The baselines we used for comparison are as follows, each selected for its unique strengths and capabilities in specific retrieval or reranking tasks:

\noindent$\bullet$ \textbf{ReasonIR}~\cite{reasonir}: ReasonIR (8B) is a reasoning-intensive retriever training on reasoning-intensive synthetic data based on LLAMA3.1-8B~\cite{llama}. This baseline is particularly adept at handling complex queries that require in-depth reasoning, making it an ideal choice as our initial retriever on the BRIGHT benchmark.

\noindent$\bullet$ \textbf{E5-Mistral-7B-Instruct}~\cite{e5-mistral}: E5-Mistral-7B-Instruct\footnote{\url{https://huggingface.co/intfloat/e5-mistral-7b-instruct}} is fine-tuned on a mixture of multilingual datasets and has some multilingual capability. We use it as our initial retriever on the R2MED benchmark.

\noindent$\bullet$ \textbf{RankT5}~\cite{rankt5}: RankT5 is a reranker that leverages the T5 architecture in a pointwise manner, optimized through a ranking loss function.

\noindent$\bullet$ \textbf{RankZephyr}~\cite{rankzephyr}: This listwise re-ranker is distilled from the powerful models GPT-3.5 and GPT-4, utilizing a sophisticated dual-phase training framework. During inference, it employs a sliding window method for listwise reranking.

\noindent$\bullet$ \textbf{Rank1}~\cite{rank1}: Rank1 functions as a pointwise reasoning reranker, distilled from the R1 model. It computes relevance scores by evaluating the probability of the final token being true or false, providing a nuanced approach to ranking that is particularly effective in scenarios requiring detailed reasoning.

\noindent$\bullet$ \textbf{Rank-R1}~\cite{rank-r1}: Rank-R1 is a setwise reasoning reranker developed through GRPO reinforcement learning. In the inference phase, it identifies the most pertinent passage from a set and employs heap sort to reorder all passages.

\noindent$\bullet$ \textbf{Rearank}~\cite{rearank}: Rearank is a listwise reasoning reranker trained using reinforcement learning.

\noindent$\bullet$ \textbf{Rank-K}~\cite{rank-k}: Rank-K represents a listwise reasoning reranker distilled from R1, trained on the MSMARCO dataset with QwQ 32B as its backbone. It utilizes a sliding window strategy during inference to rearrange all passages, which allows it to achieve a trade-off between ranking effectiveness and efficiency.

\section{Implementation Details} \label{app:implementation_details}
To help researchers better understand our approach, we have provided a comprehensive and detailed explanation of the implementation details of the Reasoning-intensive Ranking Data Synthesis and the Two-Stage Training Framework here.

\begin{table}[]
\small
\centering
\setlength{\tabcolsep}{3mm}{
\begin{tabular}{lcc}
\toprule
\multicolumn{1}{l}{\textbf{Category}} & \textbf{Dataset Name} & \textbf{Data Num} \\ \midrule
\multirow{6}{*}{\begin{tabular}[c]{@{}c@{}}Complex QA\end{tabular}} & Biology & 1700 \\
 & Earth Science & 566 \\
 & Economics & 787 \\
 & Robotics & 451 \\
 & Sustainable Living & 147 \\
 & Stackoverflow & 1741 \\ \midrule
\multicolumn{1}{l}{Coding} & Leetcode & 1633 \\ \midrule
\multicolumn{1}{l}{\multirow{2}{*}{Math}} & Math-QA & 1726 \\
\multicolumn{1}{l}{} & Math-Theorem & 1673 \\ \midrule
\multicolumn{1}{l}{Web Search} & MS MARCO & 3093 \\ \bottomrule
\end{tabular}}
\caption{The statistics of our synthesized reasoning-intensive training data (totally 13.5k) on four domains, including Complex QA, Coding, Math, and Web search.}
\label{tab:statistics}
\end{table}

\subsection{Details about Reasoning-intensive Ranking Data Synthesis}
In this part, we will introduce the construction of our training data. After constructing our data from multiple domains, we apply a self-consistency data filtering mechanism to filter low-quality training data. The parameter $\alpha$ is set as 0.4, which is determined by grid search with a step of 0.1 in the range of [0.1, 0.7] on our validation set. The statistics of final training data are shown in Table~\ref{tab:statistics}. Next, we will elaborate on the details of constructing data for each domain.

\paragraph{Complex QA}
When splitting long documents into passages, we use simple heuristics with separators such as double new lines, without making extra assumptions about the file structure. This approach allows for flexibility and adaptability to various document formats, ensuring that the passages remain coherent and contextually meaningful.

\begin{figure*}[!tb]
  \centering
  \includegraphics[width=1\linewidth]{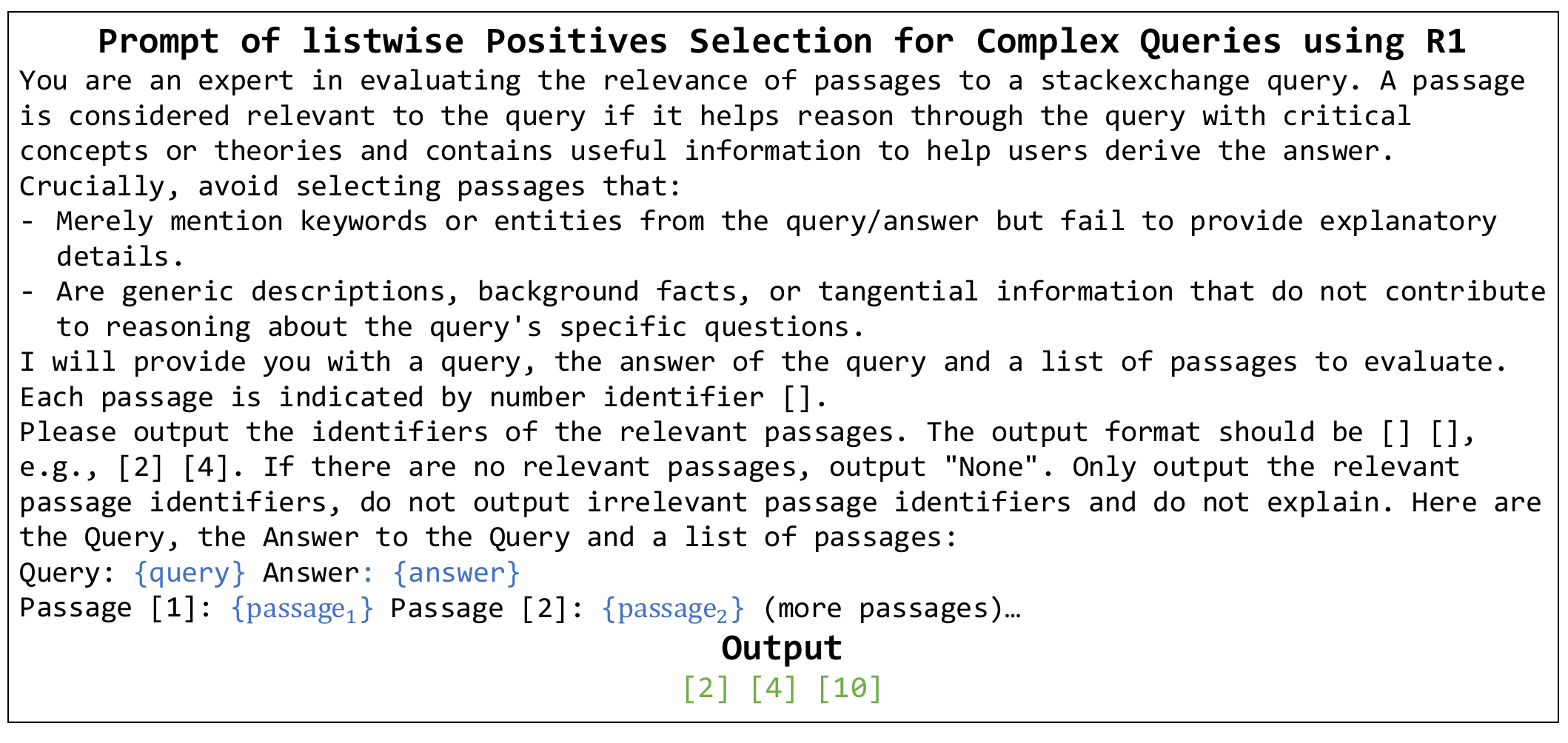}
  \caption{The prompt of listwise positive passages selection for complex queries using R1 and an example output.}
  \label{fig:positive_selection}
\end{figure*}

When using R1 to perform listwise positive passages selection, we input a list of candidate passages, the query, and its gold answer to R1, and output the IDs of positives. The prompt is shown in Figure~\ref{fig:positive_selection}.

As for the hard negative passages selection by R1, we input a list of candidate passages, the query, and its gold answer to the model, and output the IDs of hard negatives. The prompt we used is shown in Figure~\ref{fig:hardneg_selection}:

\begin{figure*}[!tb]
  \centering
  \includegraphics[width=1\linewidth]{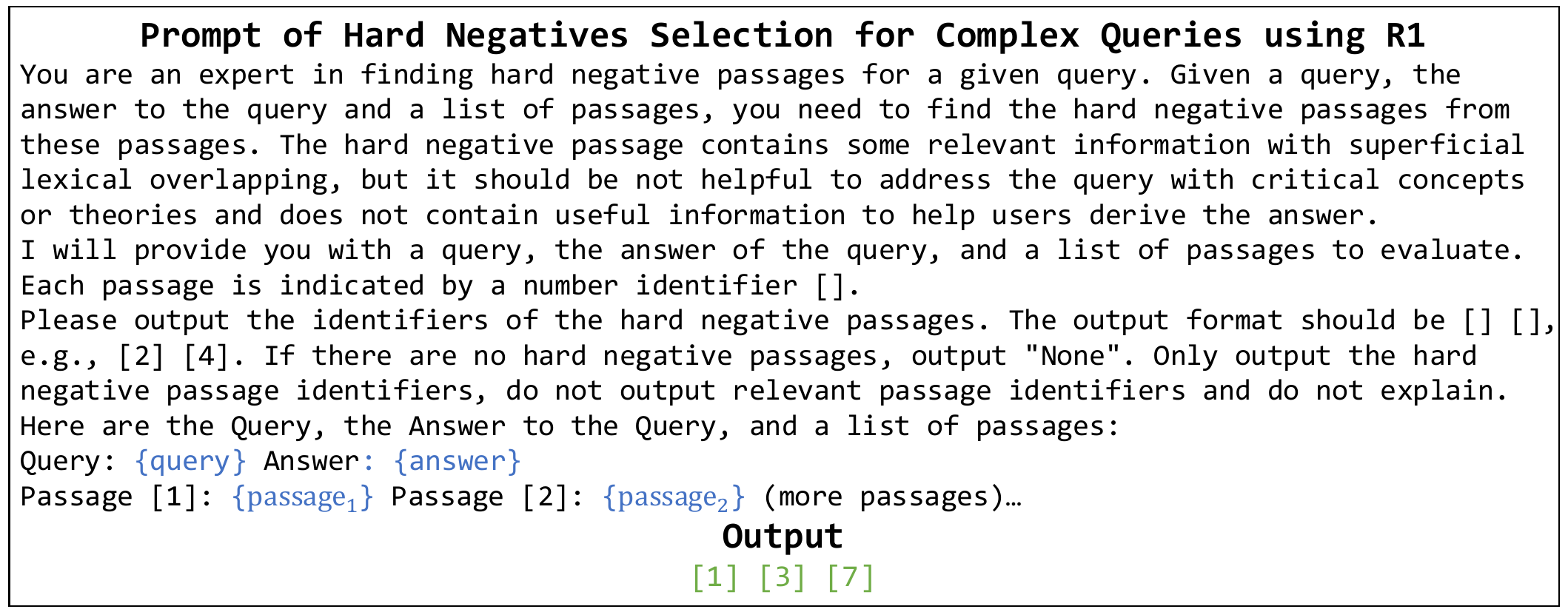}
  \caption{The prompt of hard negative passages selection for complex queries using R1 and an example output.}
  \label{fig:hardneg_selection}
\end{figure*}

After obtaining the training candidate passages, we use R1 to perform listwise ranking on these passages using the prompt in Figure~\ref{fig:r1_listwise} (the same for the subsequent data synthesis for coding, math, and web search):

\begin{figure*}[!tb]
  \centering
  \includegraphics[width=1\linewidth]{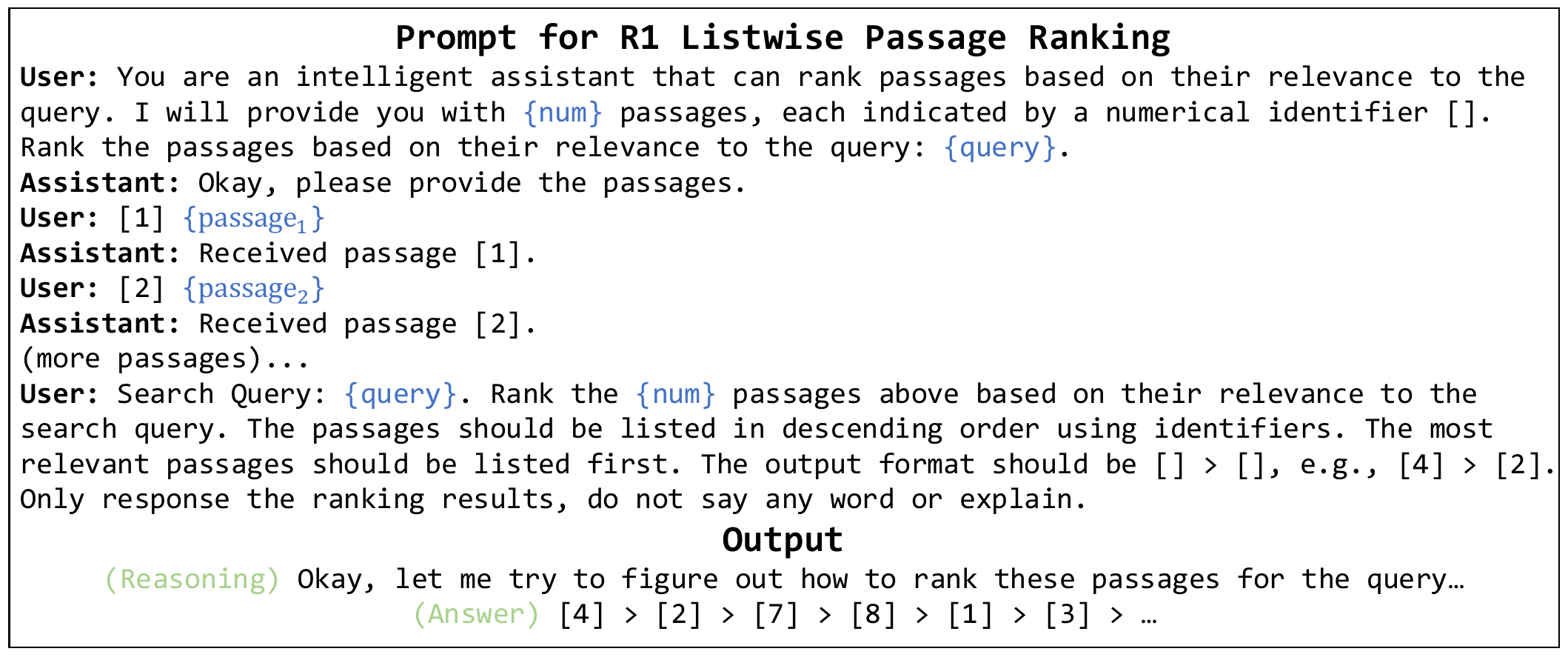}
  \caption{The prompt of R1 listwise passage ranking and an example output.}
  \label{fig:r1_listwise}
\end{figure*}

\paragraph{Coding}
As for coding queries, we apply the prompt in Figure~\ref{fig:positive_selection_coding} for mining positives:

\begin{figure*}[!tb]
  \centering
  \includegraphics[width=1\linewidth]{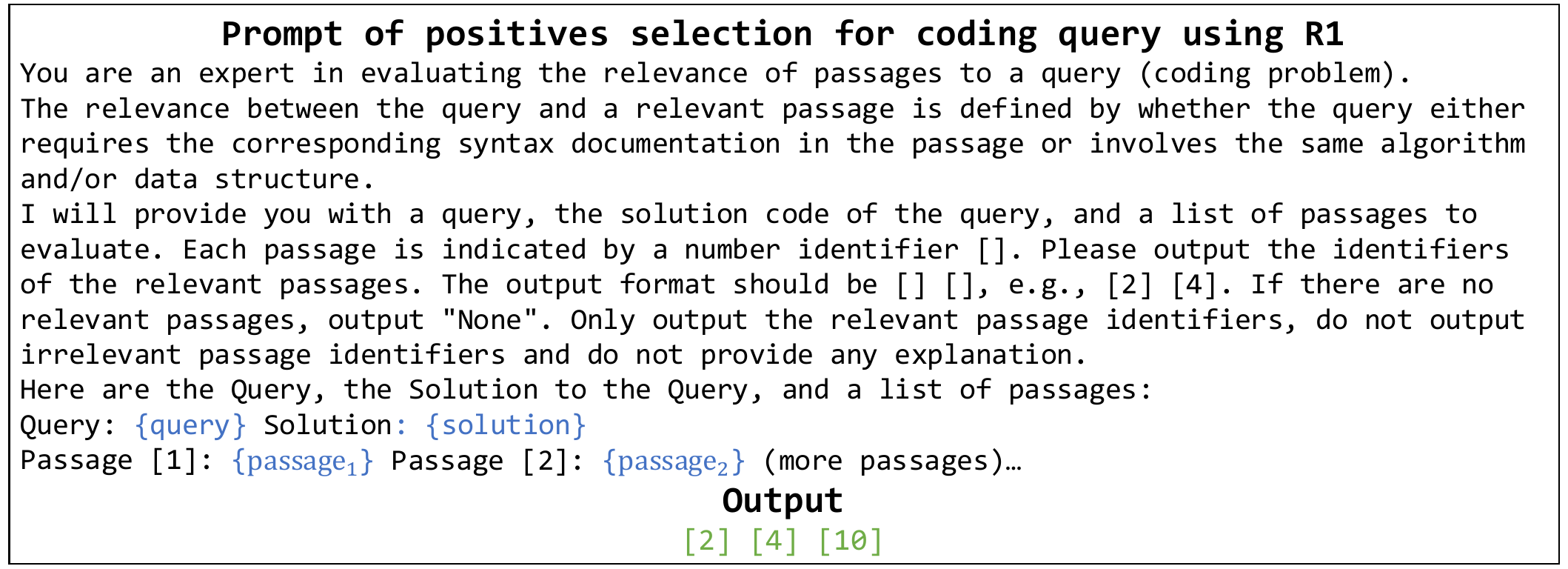}
  \caption{The prompt of positives selection for coding query using R1 and an example output.}
  \label{fig:positive_selection_coding}
\end{figure*}

\begin{figure*}[!tb]
  \centering
  \includegraphics[width=1\linewidth]{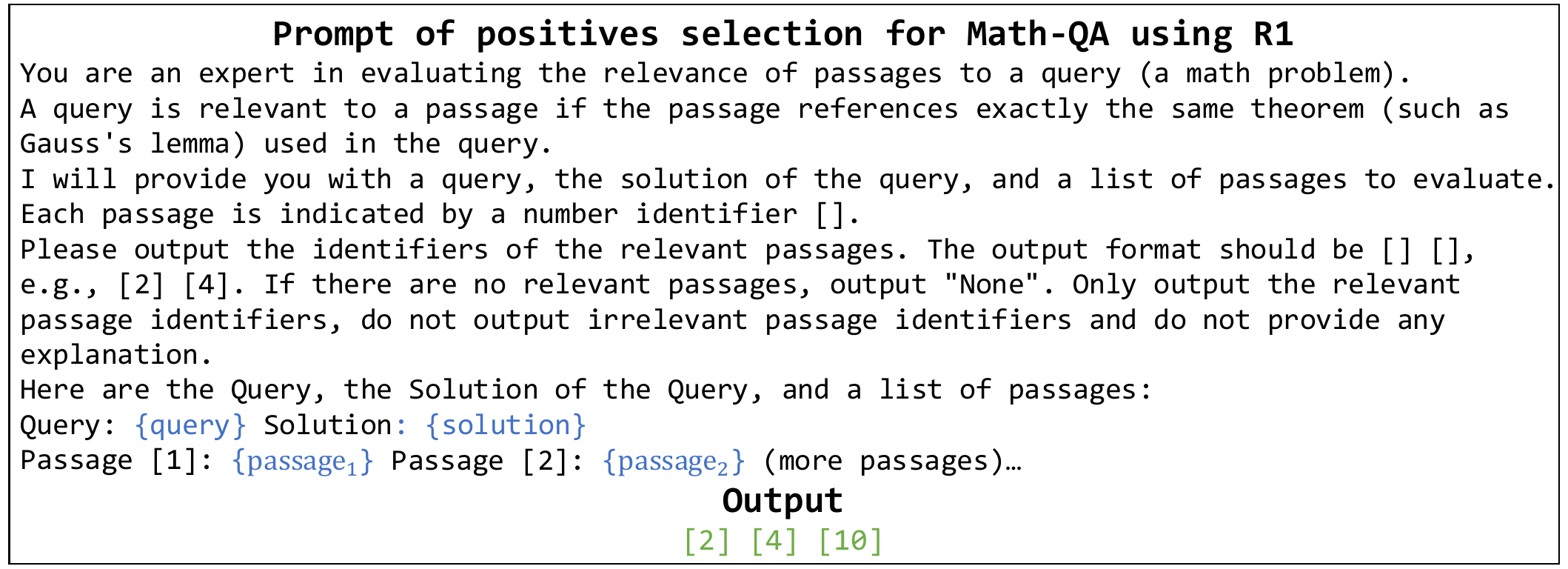}
  \caption{The prompt of positives selection for Math-QA using R1 and an example output.}
  \label{fig:positive_selection_math}
\end{figure*}

\begin{figure*}[!tb]
  \centering
  \includegraphics[width=1\linewidth]{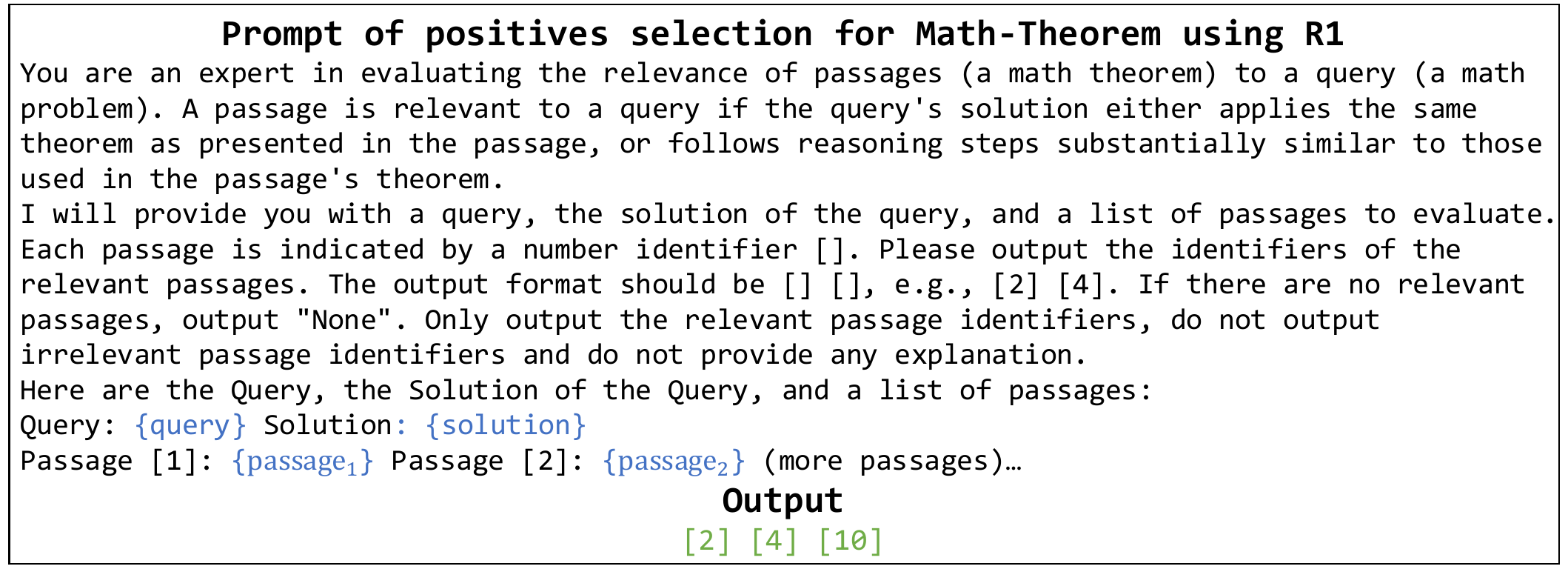}
  \caption{The prompt of positives selection for Math-Theorem using R1 and an example output.}
  \label{fig:positive_selection_math-theorem}
\end{figure*}

\paragraph{Math}
We have two different ranking tasks for math queries, including Math-QA and Math-Theorem. We use the prompts shown in Figure~\ref{fig:positive_selection_math} and Figure~\ref{fig:positive_selection_math-theorem} to select positives for Math-QA and Math-Theorem tasks based on R1, respectively.

\paragraph{Web Search}
For web search, we sample 4k training queries from the MS MARCO~\cite{MSMARCO} training set and apply the same listwise ranking prompt used for complex QA.

\subsection{Details about Two-Stage Training Framework}

\paragraph{Cold-Start SFT}
During the cold-start SFT stage, we use LlamaFactory~\cite{llamafactory} to train two different backbone LLMs: Qwen2.5-7B-Instruct\footnote{\url{https://huggingface.co/Qwen/Qwen2.5-7B-Instruct}} and Qwen2.5-32B-Instruct\footnote{\url{https://huggingface.co/Qwen/Qwen2.5-32B-Instruct}}.

As for Qwen2.5-7B-Instruct, we set the learning rate as 5e-6 and batch size per GPU as 1 with gradient accumulation steps as 8. We use DeepSpeed ZeRO-1~\cite{deepspeed} and FlashAttention2~\cite{flashattention} for our training. We apply mixed precision BF16 for training, set the maximum reasoning length as 3072, and train the model for 5 epochs.

As for Qwen2.5-32B-Instruct, we use LoRA~\cite{lora} for efficient SFT. The lora parameters rank and alpha are both set to 32. We set learning rate to 1e-4 and batch size per GPU to 1 with gradient accumulation steps of 8. We use DeepSpeed ZeRO-3~\cite{deepspeed} with mixed precision BF16 and train the model for 4 epochs.

\paragraph{Multi-view Ranking based RL} \label{app:rl}
RL has been shown in many works~\cite{tool-star, R1-Searcher++, arpo} to further enhance the performance of SFT models.
After our cold-start SFT, we further use the GRPO reinforcement learning algorithm~\cite{grpo} based on the VERL framework~\footnote{\url{https://github.com/volcengine/verl}} to train our models (\ie, ReasonRank (7B) and ReasonRank (32B)). 

The RBO reward in our multi-view ranking reward is calculated as: 
\begin{equation}
\label{eq:rbo}
\text{RBO} = (1 - p) \sum_{d=1}^{|y^{\text{list}}|}
p^{d-1} \cdot \frac{\left| y'_{1:d} \cap y^{\text{list}}_{1:d} \right|}{d}.
\end{equation}
where $y'$ and $y^{\text{list}}$ represent the rollout ranking list and gold ranking list in our listwise label, and  $p$ is a pre-defined parameter.

During GRPO training, we sample a group of output sequences $G = \{y_1, y_2, \ldots, y_G\}$ for each input $x$. Each sequence $y_i$ receives a reward $r_i$, which is then normalized within group $G$ to produce advantages $\hat{A}_i$. The token-level optimization objective is formulated as:
\begin{equation} \label{eq:grpo}
\begin{aligned}
\mathcal{J}_{\text{GRPO}}(\theta) &= \frac{1}{|G|} \sum_{i=1}^{|G|} \frac{1}{|y_i|} \sum_{t=1}^{|y_i|} \min \left(r_{i,t}(\theta) \hat{A}_{i,t}, \right. \\
&\quad \left. \text{clip} \left( r_{i,t}(\theta), 1 - \epsilon, 1 + \epsilon \right) \hat{A}_{i,t} \right) - \beta D_{\text{KL}}, \\
r_{i,t}(\theta) &= \frac{\pi_\theta(y_{i,t} \mid x, y_{i,<t})}{\pi_{\text{old}}(y_{i,t} \mid x, y_{i,<t})}, \\
D_{\text{KL}} &= D_{\text{KL}}(\pi_\theta \parallel \pi_{\text{ref}}),
\end{aligned}
\end{equation}
where $\epsilon$ and $\beta$ are hyper-parameters.

For both ReasonRank (7B) and ReasonRank (32B), the training batch size is set as 16, and the mini-batch size is set as 8. The maximum output length is set to 3072, and rollout per sample is 8. Both models are trained for 200 steps on 8 NVIDIA A800 GPUs. As for ReasonRank (7B), we use the 7B-scale SFT model as our policy model and set the learning rate to 1e-6. As for ReasonRank (32B), we use the 32B-scale SFT model as our policy model and set the learning rate to 1e-5. The lora parameters rank and alpha are both set to 32. 

\begin{figure*}[!tb]
  \centering
  \includegraphics[width=1\linewidth]{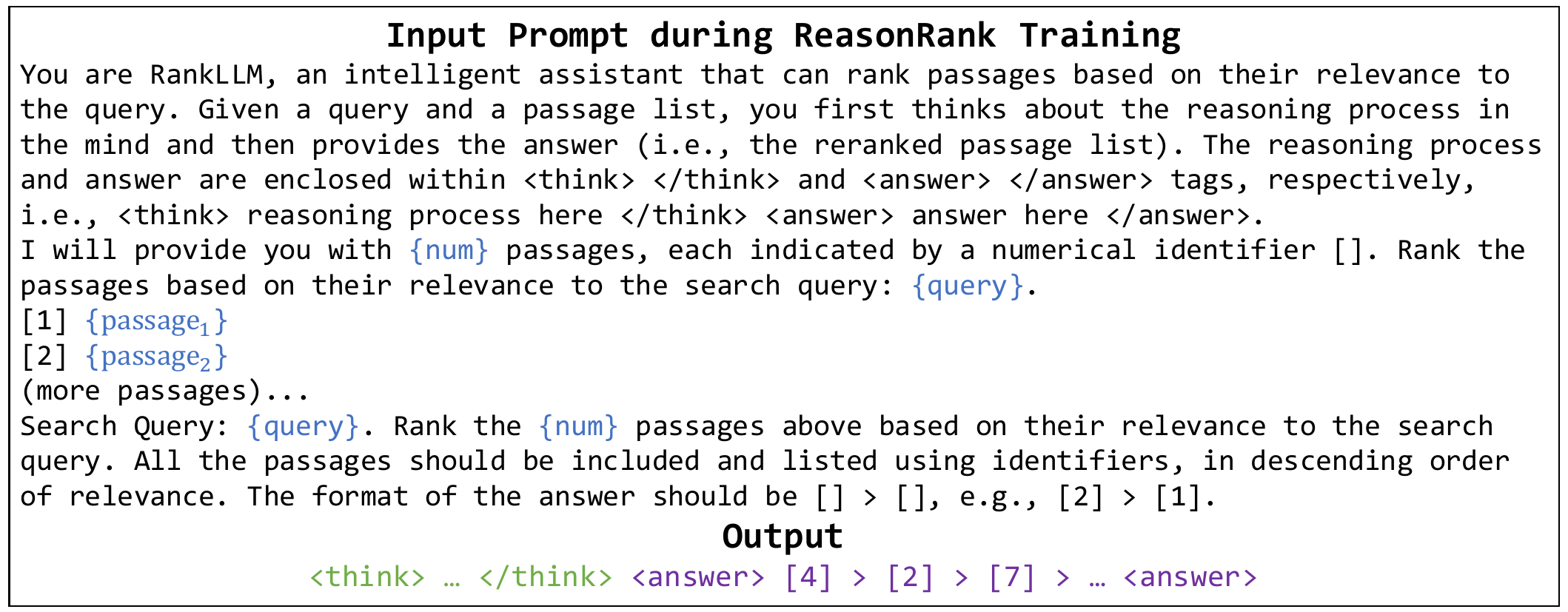}
  \caption{The input prompt of ReasonRank during training and an example output.}
  \label{fig:reasonrank_prompt}
\end{figure*}

We set the hyperparameters $\phi$ and $\gamma$ used in the multi-view ranking reward as 0.2 and 0.1, respectively, by grid search with a step size of 0.1 in the range of [0.1, 0.5]. During cold-start SFT and multi-viewranking-basedd RL, we use the input prompt in Figure~\ref{fig:reasonrank_prompt} to train ReasonRank.

\begin{table*}[!t]
\small
\centering
\setlength{\tabcolsep}{1.2mm}{
\begin{tabular}{lccccccccccccc}
\toprule
\textbf{Models} & \textbf{Econ.} & \textbf{Earth.} & \textbf{Rob.} & \textbf{Bio.} & \textbf{Psy.} & \textbf{Stack.} & \textbf{Sus.} & \textbf{Leet.} & \textbf{Pony} & \textbf{AoPS} & \textbf{TheoQ.} & \textbf{TheoT.} & \textbf{Avg.} \\ \midrule
ReasonRank (7B) & 35.05 & 47.75 & 31.22 & 56.70 & 47.77 & 32.47 & 40.87 & 23.17 & 24.95 & 7.68 & 39.49 & 41.80 & \textbf{35.74} \\ \midrule
\multicolumn{14}{l}{\textit{Training Data}} \\
only MSMARCO & 29.08 & 36.97 & 24.34 & 43.60 & 37.95 & 26.05 & 31.42 & 26.96 & 24.51 & 7.08 & 33.22 & 39.75 & 30.08 \\
\textit{w/o} Self Consistency & 33.96 & 50.35 & 27.94 & 55.48 & 46.63 & 31.96 & 40.17 & 22.50 & 18.29 & 7.32 & 35.16 & 40.67 & 34.20 \\ \midrule
\multicolumn{14}{l}{\textit{Training Approach}} \\
\textit{w/o} SFT (only RL) & 27.44 & 35.23 & 23.73 & 50.36 & 34.48 & 25.76 & 29.62 & 18.59 & 23.51 & 6.78 & 34.46 & 34.28 & 28.69 \\
\textit{w/o} RL (only SFT) & 30.47 & 44.83 & 27.42 & 50.53 & 43.10 & 30.38 & 38.79 & 25.11 & 22.46 & 7.06 & 37.53 & 40.12 & 33.15 \\
\textit{w/o} $R^m$ & 30.86 & 47.14 & 29.75 & 59.6 & 44.40 & 35.04 & 36.87 & 20.47 & 20.80 & 8.23 & 37.52 & 39.66 & 34.20 \\
non-reasoning SFT & 26.70 & 45.30 & 27.89 & 51.48 & 43.65 & 29.72 & 41.44 & 25.60 & 17.64 & 5.39 & 38.31 & 42.4 & 32.96 \\ \bottomrule
\end{tabular}}
\caption{The performance (NDCG@10) of ablated models on each BRIGHT dataset.}
\label{tab:ablation_detailed}
\end{table*}

\begin{table*}[!t]
\small
\centering
\setlength{\tabcolsep}{2.9mm}{
\begin{tabular}{lcccccccc}
\toprule
\textbf{Models} & \textbf{Covid} & \textbf{DBPedia} & \textbf{SciFact} & \textbf{NFCorpus} & \textbf{Signal} & \textbf{Robust04} & \textbf{News} & \textbf{Avg.} \\ \midrule
BM25 & 59.47 & 31.80 & 67.89 & 33.75 & 33.04 & 40.70 & 39.52 & 43.74 \\ \midrule
\textit{\textbf{Non-reasoning reranker}} &  &  &  &  &  &  &  &  \\
RankT5 (3B) & 80.19 & 44.85 & 74.60 & 37.40 & 31.73 & 51.45 & 49.11 & 52.76 \\
RankZephyr (7B) & 82.92 & 44.42 & 75.42 & 38.26 & 31.41 & 53.73 & \textbf{52.80} & 54.14 \\ \midrule
\textit{\textbf{Reasoning reranker}} &  &  &  &  &  &  &  &  \\
Rank-R1 (7B) & 83.71 & 42.27 & 72.16 & 38.94 & \textbf{33.08} & 54.46 & 50.60 & 53.60 \\
Rank-R1 (14B) & \textbf{84.63} & 44.05 & 75.96 & 38.58 & 32.95 & 56.91 & 49.20 & 54.61 \\
Rank1 (7B) & 79.04 & 35.79 & 73.32 & 37.52 & 25.41 & 57.11 & 47.67 & 50.84 \\
Rank1 (32B) & 80.59 & 34.79 & 74.78 & 37.29 & 25.58 & 58.30 & 45.57 & 50.99 \\
Rearank (7B) & 81.28 & 44.99 & 75.02 & 38.83 & 32.69 & 56.16 & 51.04 & 54.29 \\
Rank-K (32B) & 74.76 & 38.29 & 60.83 & 36.48 & 32.64 & 49.79 & 45.59 & 48.34 \\ 
\rowcolor{myblue}ReasonRank (7B) & 82.01 & \textbf{46.03} & 75.55 & 39.60 & 31.36 & 55.40 & 50.50 & 54.35 \\
\rowcolor{myblue}ReasonRank (32B) & 83.16 & 45.65 & \textbf{77.24} & \textbf{40.04} & 31.07 & \textbf{58.67} & 52.24 & \textbf{55.44} \\ \bottomrule
\end{tabular}}
\caption{The results (NDCG@10) on 7 BEIR datasets. All models rerank BM25-retrieved top-100 passages. The best reranker is highlighted in \textbf{bold}.}
\label{tab:beir_detailed}
\end{table*}

\begin{table*}[!t]
\small
\centering
\setlength{\tabcolsep}{1.1mm}{
\begin{tabular}{lccccccccccccc}
\toprule
\textbf{Models} & \textbf{Econ.} & \textbf{Earth.} & \textbf{Rob.} & \textbf{Bio.} & \textbf{Psy.} & \textbf{Stack.} & \textbf{Sus.} & \textbf{Leet.} & \textbf{Pony} & \textbf{AoPS} & \textbf{TheoQ.} & \textbf{TheoT.} & \textbf{Avg.} \\ \midrule
ReasonIR (8B) & 32.65 & 43.00 & 20.82 & 43.49 & 39.57 & 30.96 & 27.34 & 31.69 & 19.55 & 7.37 & 33.93 & 36.68 & 30.59 \\ \midrule
rankT5 (3B) & 11.35 & 22.11 & 10.94 & 13.62 & 11.40 & 11.35 & 15.96 & 27.45 & {\ul 38.05} & \textbf{9.24} & 18.27 & 9.46 & 16.60 \\
RankZepyer (7B) & 19.87 & 17.36 & 12.35 & 34.90 & 24.72 & 13.35 & 22.34 & {\ul 29.29} & 32.37 & 6.05 & 28.98 & 30.07 & 22.64 \\
Rank-R1 (7B) & 20.19 & 27.91 & 18.13 & 36.70 & 30.22 & 11.32 & 29.28 & 17.13 & 9.35 & 3.23 & 14.43 & 28.98 & 20.57 \\
Rank-R1 (14B) & 27.39 & 38.73 & 23.11 & 44.45 & 37.10 & 27.82 & 36.77 & 21.27 & 19.23 & 8.80 & 31.66 & 39.53 & 29.66 \\
Rank1 (7B) & 25.34 & 38.91 & 16.77 & 39.91 & 35.32 & 24.81 & 33.47 & 12.71 & 28.14 & 2.58 & 30.67 & 38.18 & 27.23 \\
Rank1 (32B) & 25.43 & 37.97 & 17.14 & 42.43 & 34.85 & 23.82 & 31.18 & 12.20 & \textbf{40.98} & 4.79 & 29.28 & 39.95 & 28.34 \\
Rearank (7B) & 30.65 & 40.27 & 27.16 & 46.94 & 40.77 & 26.12 & 36.26 & \textbf{30.55} & 22.84 & 7.31 & 32.28 & 39.86 & 31.75 \\
Rank-K (32B) & 30.09 & 39.78 & 26.60 & 50.64 & 43.45 & 29.87 & 35.17 & 27.23 & 22.80 & 7.61 & 37.06 & 41.00 & 32.61 \\ \midrule
\multicolumn{14}{l}{\textit{Using Qwen2.5-7B-Instruct as the backbone}} \\
\rowcolor{myblue}$\text{ReasonRank}_\text{Only-SFT}$ (7B) & 30.47 & {\ul 44.83} & 27.42 & 50.53 & 43.10 & 30.38 & 38.79 & 25.11 & 22.46 & 7.06 & 37.53 & 40.12 & 33.15 \\
\rowcolor{myblue}ReasonRank (7B) & \textbf{35.05} & \textbf{47.75} & {\ul 31.22} & \textbf{56.70} & 47.77 & {\ul 32.47} & 40.87 & 23.17 & 24.95 & 7.68 & \textbf{39.49} & 41.80 & {\ul 35.74} \\ \midrule
\multicolumn{14}{l}{\textit{Using Meta-Llama-3.1-8B-Instruct as the backbone}} \\
\rowcolor{myblue}$\text{ReasonRank}_\text{Only-SFT}$ (8B) & 30.58 & 41.81 & 27.58 & 50.48 & 48.24 & 27.49 & 39.61 & 22.52 & 25.40 & 7.45 & 35.46 & 35.54 & 32.68 \\
\rowcolor{myblue}ReasonRank (8B) & 34.47 & 40.60 & 27.46 & 53.44 & \textbf{51.86} & 29.52 & 43.58 & 22.40 & 23.79 & 7.69 & 34.44 & 39.80 & 34.08 \\ \midrule
\multicolumn{14}{l}{\textit{Using Qwen3-8B as the backbone}} \\
\rowcolor{myblue}$\text{ReasonRank}_\text{Only-SFT}$ (8B) & 34.67 & 44.18 & \textbf{31.73} & 53.31 & 45.55 & 30.58 & \textbf{44.12} & 24.00 & 25.37 & 8.11 & 37.50 & {\ul 43.60} & 35.22 \\
\rowcolor{myblue}ReasonRank (8B) & {\ul 34.71} & 44.23 & 29.46 & {\ul 55.16} & {\ul 50.10} & \textbf{32.50} & {\ul 43.96} & 24.23 & 28.44 & {\ul 8.84} & {\ul 37.82} & \textbf{44.19} & \textbf{36.14} \\ \bottomrule
\end{tabular}}
\caption{The performance of ReasonRank (7B) based on different backbones.}
\label{tab:different_backbones}
\end{table*}

\section{More Experiments for ReasonRank}

\subsection{Detailed Results of Ablation Study} \label{app:ablation}
We show the results of our ablation study based on ReasonRank (7B) on each BRIGHT dataset in Table~\ref{tab:ablation_detailed}. The ablated models show performance decline on almost all datasets, proving the effectiveness of the corresponding component.

\subsection{Detailed Results on BEIR Benchmark} 
In this part, we present the comparison results on 7 datasets of the traditional IR benchmark BEIR~\cite{beir}. We use BM25 as our retriever and rerank the top-100 retrieved passages. We also add rankT5 (3B), Rank-R1 (7B), and Rank1 (7B) baselines for a more comprehensive comparison. The results are shown in Table~\ref{tab:beir_detailed}. 

\begin{figure*}[!t]
        \centering
	\includegraphics[width=1\linewidth]{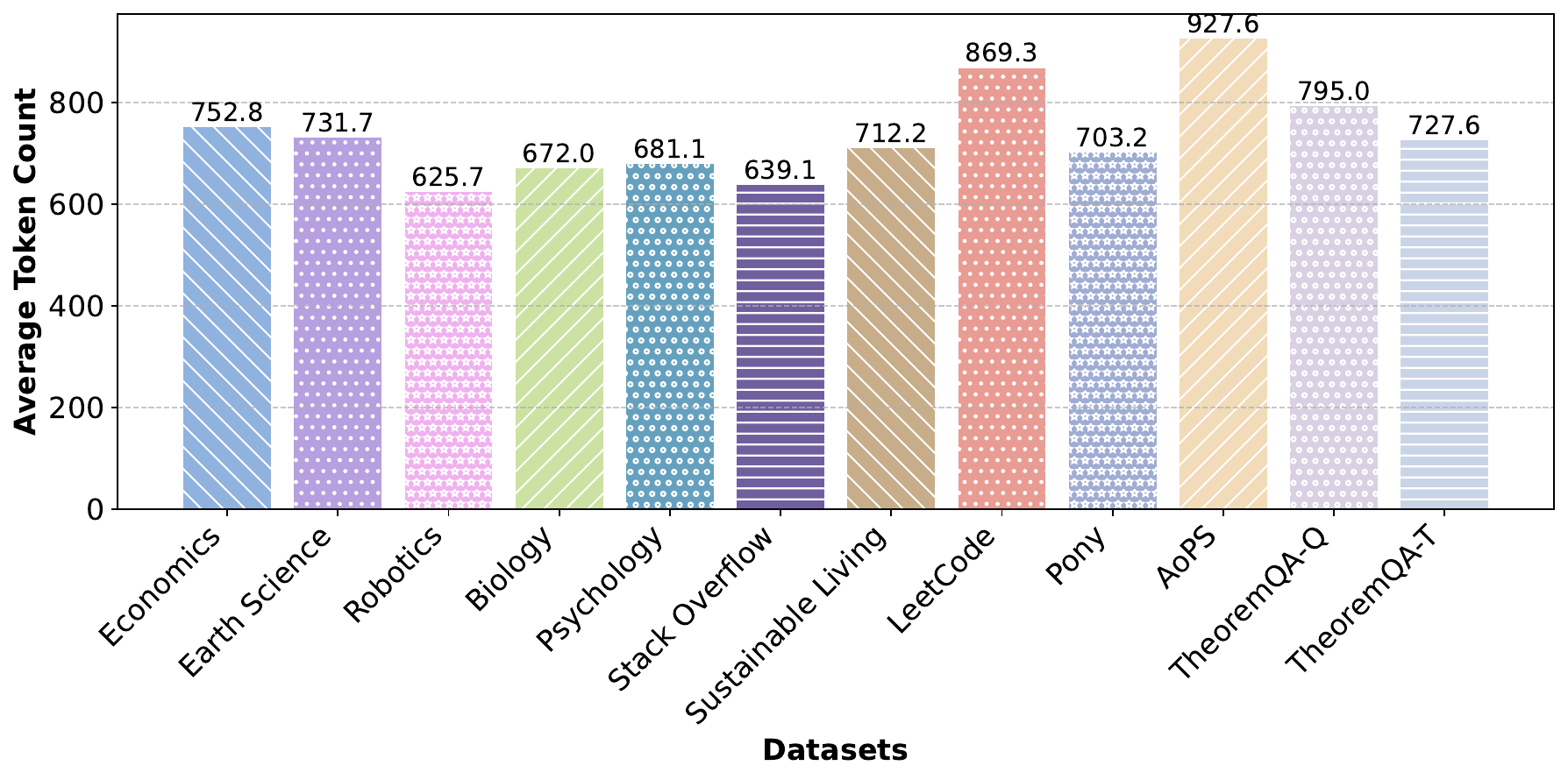}
	\caption{The reasoning length of ReasonRank (7B) on BRIGHT.}
	\label{fig:reasoning_len}
\end{figure*}

\subsection{Different Backbones} \label{subsec:different_backbones}
To demonstrate the generalizability and effectiveness of our proposed ReasonRank framework, we conduct additional experiments using different backbone LLMs. In addition to Qwen2.5-7B-Instruct utilized in our main experiments, we further adopt a reasoning LLM Qwen3-8B and Meta-Llama-3.1-8B-Instruct, which represents a different model family from Qwen. The results are shown in Table~\ref{tab:different_backbones}. We also report the performance of ReasonRank, which is only trained using supervised fine-tuning (SFT), denoted as $\text{ReasonRank}_\text{Only-SFT}$. 

As shown in Table~\ref{tab:different_backbones}, our ReasonRank consistently and significantly outperforms all baseline models across all backbones, validating the effectiveness of our synthetic data generation pipeline and the two-stage training strategy. Moreover, ReasonRank exhibits obvious performance improvements over $\text{ReasonRank}_\text{Only-SFT}$ on Avg. metric, which further substantiates the benefit of our multi-view ranking-based RL approach. Notably, when built upon Qwen3-8B, ReasonRank achieves the best overall average score of 36.14\%, suggesting that the inherent strong reasoning capability of Qwen3 may further enhance the ranking performance of our framework.

\subsection{Analysis of Reasoning Length}
In this section, we explore the average length of reasoning chains using ReasonRank across various BRIGHT datasets. We select ReasonRank (7B) and rerank the top-20 passages retrieved by ReasonIR. The average reasoning length for queries in each dataset is shown in Figure~\ref{fig:reasoning_len}. From the results, we observe that the reasoning length of coding tasks (\ie, Leetcode and Pony) and theorem tasks (\eg, AoPS, TheoremQA-Q and TheoremQA-T) tend to be longer than StackExchange tasks (\eg, Robotics and Stack Overflow). This may be because coding queries and theorem queries are more complex and challenging compared to StackExchange queries, thus requiring longer reasoning to understand queries and rerank the passages.

\begin{table*}[!t]
\small
\centering
\setlength{\tabcolsep}{1.2mm}{
\begin{tabular}{lccccccccccccc}
\toprule
\textbf{Models} & \textbf{Econ.} & \textbf{Earth.} & \textbf{Rob.} & \textbf{Bio.} & \textbf{Psy.} & \textbf{Stack.} & \textbf{Sus.} & \textbf{Leet.} & \textbf{Pony} & \textbf{AoPS} & \textbf{TheoQ.} & \textbf{TheoT.} & \textbf{Avg.} \\ \midrule
ReasonIR (8B) & 32.65 & 43.00 & 20.82 & 43.49 & 39.57 & 30.96 & 27.34 & 31.69 & 19.55 & 7.37 & 33.93 & 36.68 & 30.59 \\ \midrule
ReasonRank (7B) & 31.32 & 43.18 & 27.07 & 50.06 & 45.02 & 33.46 & 37.25 & 24.14 & \textbf{22.93} & 8.23 & \textbf{35.39} & 40.69 & 33.23 \\
\rowcolor{myblue}ReasonRank (32B) & \textbf{33.98} & 42.30 & \textbf{28.66} & \textbf{51.97} & 47.45 & 36.31 & \textbf{38.51} & 23.15 & 22.48 & \textbf{8.59} & 34.51 & 42.27 & \textbf{34.18} \\
DeepSeek-R1 (671B) & 33.75 & \textbf{45.64} & 26.91 & 50.34 & \textbf{49.47} & \textbf{37.22} & 36.92 & \textbf{25.74} & 21.08 & 6.81 & 33.30 & \textbf{42.42} & 34.13 \\ \bottomrule
\end{tabular}}
\caption{The comparison between ReasonRank and DeepSeek-R1. All models rerank ReasonIR-retrieved top-20 passages. We use GPT-4 rewritten queries during retrieval and use the original query for reranking. The best reranker is marked in bold.}
\label{tab:compare_with_teacher}
\end{table*}

\begin{table*}[!t]
\small
\centering
\setlength{\tabcolsep}{0.9mm}{
\begin{tabular}{lccccccccccccc}
\toprule
\textbf{Models} & \textbf{Econ.} & \textbf{Earth.} & \textbf{Rob.} & \textbf{Bio.} & \textbf{Psy.} & \textbf{Stack.} & \textbf{Sus.} & \textbf{Leet.} & \textbf{Pony} & \textbf{AoPS} & \textbf{TheoQ.} & \textbf{TheoT.} & \textbf{Avg.} \\ \midrule
ReasonIR (8B) & 32.65 & 43.00 & 20.82 & 43.49 & 39.57 & 30.96 & 27.34 & 31.69 & 19.55 & 7.37 & 33.93 & 36.68 & 30.59 \\
ReasonRank (32B) & 36.64 & 48.90 & 33.88 & 58.17 & 53.27 & 38.68 & 45.97 & 25.78 & 20.93 & 9.41 & 38.46 & 46.28 & 38.03 \\ \midrule
RaDeR + BM25 (Hybrid) & 30.85 & 56.08 & 24.13 & 52.66 & 46.69 & 34.93 & 32.22 & 31.91 & 25.87 & 11.98 & 37.73 & 43.43 & 35.71 \\
(1) ReasonRank (32B) & 37.67 & 50.61 & 35.44 & 59.04 & 53.06 & 39.04 & 45.75 & 22.64 & 25.30 & 11.41 & 41.27 & 47.73 & 39.08 \\
\rowcolor{myblue}(2) ReasonRank' (32B) & 36.70 & 55.53 & 35.69 & 62.72 & 54.64 & 38.03 & 44.81 & 29.46 & 25.56 & 14.38 & 41.99 & 50.06 & \textbf{40.80} \\ \bottomrule
\end{tabular}}
\caption{Further ranking enhancement on BRIGHT. Both retrievers ``ReasonIR (8B)'' and ``RaDeR + BM25 (Hybrid)'' use GPT4-rewritten queries for retrieval.}
\label{tab:further_enhancement_all}
\end{table*}

\subsection{Comparison with Teacher DeepSeek-R1}
During our reasoning-intensive training data synthesis, we use DeepSeek-R1 to mine positive and negative passages based on the gold answer to obtain pointwise training labels. Theoretically, this should generate better labels compared to pure teacher distillation (which does not use gold answers and is based only on the teacher), potentially allowing the trained model to outperform DeepSeek-R1. To verify this, we compare the ranking performance of our trained ReasonRank (7B, 32B) with DeepSeek-R1 on BRIGHT. Considering the high time and API cost of reranking with DeepSeek-R1, we choose to rank the top-20 (instead of top-100) passages retrieved by ReasonIR. The ranking prompt used by R1 is the same as the prompt used when generating listwise labels based on R1. The results are shown in Table~\ref{tab:compare_with_teacher}. From the results, we can see that ReasonRank (32B) achieves comparable performance to DeepSeek-R1 (34.18 vs 34.13) and surpasses DeepSeek-R1 on 7 datasets (Economics, Robotics, Biology, Sustainable Living, Pony, AoPS, and TheoremQA-Q), proving the superiority of training with our synthesized data compared to pure teacher distillation.


\subsection{Further Ranking Enhancement}
Although our ReasonRank has already achieved superior results compared to existing rerankers, we believe its ranking performance can be further improved by adjusting some ranking settings. Previous studies~\cite{rankvicuna, rankzephyr} have revealed that the quality of retrieval results has an impact on ranking performance: better retrieval results can lead to better ranking performance to some extent. In this section, we used the retrieval results provided by RaDeR~\cite{rader}, which combines the retrieval results of the dense retriever RaDeR and the sparse retriever BM25 and achieves higher retrieval quality than ReasonIR. We apply ReasonRank (32B) as our reranker to rerank the top-100 retrieved passages. As shown in Figure~\ref{tab:further_enhancement_all} (1), with better retrieval results, the ranking performance further improved to 39.08 (1 point higher than the ranking results based on ReasonIR), which aligns with the previous conclusion. Furthermore, previous work~\cite{rankgpt} pointed out that the size and stride of the sliding window also affect listwise ranking performance. In this section, different from the original settings with a window size of 20 and a stride of 10, we experimented with new sliding window parameters with a window size of 10 and a stride of 5 (denoted as $\text{ReasonRank' (32B)}$). The results in Table~\ref{tab:further_enhancement_all} (2) show that this change brought about a 1.7-point improvement. This may be because a smaller window size can reduce the difficulty of modeling context and ranking in listwise ranking, thereby improving performance. \textbf{Remarkably, with an average score of 40.6, ReasonRank (32B) has achieved SOTA performance on the BRIGHT leaderboard on August 9, 2025.}

\section{Case Studies of ReasonRank (7B)}
To better understand the reasoning process of our ReasonRank, we present the reasoning chains of our ReasonRank (7B) on several BRIGHT datasets, including Earth Science (Table~\ref{tab:case_earth_science}), Biology (Table~\ref{tab:case_biology}), Robotics (Table~\ref{tab:robotics}), LeetCode (Table~\ref{tab:LeetCode}), AoPS (Table~\ref{tab:AoPS}), and TheoremQA-Theorems (Table~\ref{tab:theoremqa-t}). We use the ReasonIR-retrieved top-20 passages as the candidate passages to rerank.

\section{Use of AI Assistants}
We use ChatGPT to improve the presentations of this paper.\footnote{\url{https://chatgpt.com/}}

\clearpage

\begin{table*}[!t]
\centering
\caption{Case on Earth Science. The model first analyzes the search query and considers the criteria that make a passage relevant. Then, it examines some relevant passages and compares them to the criteria. Finally, it quickly goes through the remaining passages and determines the final ranking.}
\fontsize{9pt}{11pt}\selectfont
\begin{tabular}{p{0.98\linewidth}}
\midrule
\rowcolor{gray!20}\textbf{Query} \\
\midrule
The line on the ground at the Greenwich Observatory is itself on a tectonic plate that is moving, so is the prime meridian still 0 degrees, 0 minutes, 0 seconds?

If not - what are the implications (if any)? For example, if my GPS says I am 100 km west of Greenwich Observatory, am I 100 km west of 0 0 0 or 100 km west of the line on the ground? Who (and what) decides this kind of thing?

Curious punter, not a scientist....
\\
\midrule

\rowcolor{gray!20}\textbf{ReasonRank Output}\\
\midrule
\green{<think>}
Okay, let me try to figure out how to rank these passages for the user's query. The main question is whether the prime meridian at Greenwich is still exactly 0° 0' 0" because the tectonic plate is moving, and what that means for GPS coordinates. The user is also curious about who decides this.

First, I need to identify which passages address the movement of the prime meridian due to tectonic shifts and how that affects GPS. The query mentions tectonic plates moving, so passages explaining that the physical line has moved are key. Also, implications like GPS readings and reference frames are important.

Passage [1] seems very relevant. It talks about the Airy Transit Circle at Greenwich and how its longitude is now 0°00'05.3'' West in ITRF and WGS84. It mentions the offset of 102m and discusses astronomical vs geodetic coordinates. This directly answers if the prime meridian has moved and explains GPS references.

Passage [3] also discusses the offset and explains it through deflection of the vertical and geodetic coordinates. It addresses GPS receivers showing east of zero, which relates to the user's example. So this is probably high relevance.

Passage [8] talks about Earth's drift due to mantle movements, ice melting, etc., and mentions the pole drift. While not directly about Greenwich, it explains why tectonic/mass movements cause shifts, which ties into the prime meridian's position.

Passage [9] discusses polar motion, which is about Earth's rotational axis, but the user's query is about longitude. However, it might touch on reference frames, so could be somewhat relevant.

Passage [10] explains why Greenwich was chosen in 1884 and mentions tectonic maps using British ones. It's historical but might not address the current movement issue.

Passage [11] is titled "Why the Greenwich meridian moved" – that sounds directly relevant. It likely explains the reasons for the shift, so high priority.

Passage [12] mentions the prime meridian through Greenwich and longitude measurement. It's more about definition than movement, so maybe less relevant.

Passage [14] discusses geographical coordinates and WGS, which relates to GPS references. Explains how modern systems differ from simple models.

Passage [7] and [18] seem to be about the 1884 conference and references to the meridian movement, so they might explain the decision-making body (BIH, IERS), which answers who decides.

Passage [2] mentions BIH Terrestrial System and geodetic references, which ties into reference frames and who decides.

Passage [4] and [5] talk about Earth's equator and axis changes, but not directly about the prime meridian's movement.

Passage [16] is about geomagnetic poles, which is a different type of movement, so less relevant.

Passage [19] and [20] explain longitude but not the tectonic aspect.

So the top ones would be [1], [3], [11], [8], [7], [18], [2], [14], [10], then others. Need to check each again.

[1] is definitely top as it directly answers the movement and GPS. [3] explains the offset and geodetic systems. [11] is about why it moved, so very relevant. [8] explains Earth drift causes. [7] and [18] mention BIH and IERS, which decide references. [2] talks about reference frames. [14] about WGS. [10] historical context. Others less so.

So the order might be [1] > [3] > [11] > [8] > [7] > [18] > [2] > [14] > [10] > [9] > [4] > [5] > [16] > [12] > [17] > [20] > [13] > [6] > [19] > [15]
\green{</think>}

\purple{<answer>} 

[1] > [3] > [11] > [8] > [7] > [18] > [2] > [14] > [10] > [9] > [4] > [5] > [16] > [12] > [17] > [20] > [13] > [6] > [19] > [15]

\purple{</answer>}
\\
\bottomrule
\end{tabular}
\label{tab:case_earth_science} 
\end{table*}

\begin{table*}[!t]
\centering
\caption{Case on Biology. The model first analyzes the search query and considers the criteria that make a passage relevant. Then, it examines some relevant passages and compares them to the criteria. Finally, it quickly goes through the remaining passages and determines the final ranking.}
\fontsize{9pt}{11pt}\selectfont
\begin{tabular}{p{0.98\linewidth}}
\midrule
\rowcolor{gray!20}\textbf{Query} \\
\midrule
Why does hair turn grey or white, and why does it happen later for some?
The question is pretty simple: what is happening molecularly when hair turns grey or white? I would imagine that it is due to the lack of a particular compound. I'm also interested in why some people's onset of grey hair happens much later than others.
\\
\midrule

\rowcolor{gray!20}\textbf{ReasonRank Output}\\
\midrule
\green{<think>}
Okay, let me try to figure out how to rank these passages. The user wants to know why hair turns gray or white molecularly and why some people gray later. So I need to focus on passages that explain the biological mechanisms, like lack of a compound, melanocyte stem cells, oxidative stress, telomeres, etc.

Passage [1] talks about melanocyte stem cells and Bcl2, which affects survival. That's relevant. Passage [2] discusses mouse models and factors like oxidative damage and telomerase, so that's good. Passage [4] mentions melanocyte stem cells and homeostasis, which ties into the molecular process. Passage [5] is an abstract about melanocyte stem cells and aging, so key. Passage [6] details melanocyte compartments and stem cells, explaining the process. Passage [8] covers melanocyte stem cell properties and bulge niche, which is crucial. Passage [15] includes DNA repair and telomere theories, which relate to aging and molecular causes. Passage [16] mentions melanocyte stem cell maintenance in the niche, so that's relevant. 

Passages [3], [7], [10], [11], [12], [18], [20] have some info but less directly on the molecular mechanisms. Others like [9], [13], [14], [17] are more about hair care or demographics, so less relevant. 

So the top ones are those explaining stem cell depletion, oxidative stress, telomeres, and niche environment. Then others that support those points. The care-related ones are less important here.
\green{</think>}

\purple{<answer>} 

[4] > [1] > [5] > [6] > [8] > [2] > [15] > [16] > [10] > [3] > [7] > [12] > [20] > [18] > [9] > [11] > [13] > [14] > [17] > [19]

\purple{</answer>}
\\
\bottomrule
\end{tabular}
\label{tab:case_biology} 
\end{table*}

\begin{table*}[!t]
\centering
\caption{Case study on Robotics. The model successfully rank the passage with relatively short reasoning length.}
\fontsize{9pt}{11pt}\selectfont
\begin{tabular}{p{0.98\linewidth}}
\midrule
\rowcolor{gray!20}\textbf{Query} \\
\midrule
ROS2: Adding parameters to YAML launch file

How do I use a YAML launch file to pass parameters to a node? The tutorial covers passing parameters using a python launch file but not YAML.
\\

\midrule

\rowcolor{gray!20}\textbf{ReasonRank Output}\\
\midrule
\green{<think>}
Okay, let me try to figure out how to rank these passages for the user's query about using YAML launch files to pass parameters in ROS2. The user mentioned that the tutorial covers Python launch files but not YAML, so they're specifically looking for YAML examples.

First, I'll go through each passage to see which ones directly address YAML launch files and parameters.

Passage [2] has a YAML config example in the launch file, mentioning \texttt{turtlesim\_custom\_params.yaml}, which seems very relevant. Passage [3] talks about the \texttt{<rosparam>} tag used in launch files with YAML, which is key here. Passage [8] also explains the \texttt{<rosparam>} tag in the context of YAML files. Passages [13]--[18] all mention \texttt{rosparam} and using YAML with \texttt{roslaunch}, which is related. Passage [19] discusses debugging with YAML files. Passage [6] gives an example of YAML parameters for Nav2, which is a good example. Passage [5] mentions YAML config files for parameters. Passage [1] also talks about YAML config files for parameters. Passages [10] and [11] are about parameter defaults and loading from YAML.

The others, like [4], [7], [12], and [20], are either about Python or C++ or not directly answering the YAML question.

So the most relevant are [2], [3], [8], then [13]-[18], [6], [5], [19], [1], followed by others. Need to make sure the order prioritizes direct YAML examples and explanations of the `<rosparam>` tag.
 \green{</think>}

\purple{<answer>} 

[2] > [3] > [8] > [13] > [14] > [15] > [16] > [17] > [18] > [6] > [5] > [19] > [1] > [10] > [11] > [9] > [4] > [7] > [12] > [20]

\purple{</answer>}
\\
\bottomrule
\end{tabular}
\label{tab:robotics} 
\end{table*}

\begin{table*}[!t]
\centering
\caption{Case on LeetCode. The model rethinks its initial ranking and repeatedly verifies and compares the top-ranked passages..}
\fontsize{9pt}{11pt}\selectfont
\begin{tabular}{p{0.98\linewidth}}
\midrule
\rowcolor{gray!20}\textbf{Query} \\
\midrule
Given an `m x n` integers `matrix`, return the length of the longest increasing path in `matrix`.

From each cell, you can either move in four directions: left, right, up, or down. You **may not** move **diagonally** or move **outside the boundary** (i.e., wrap-around is not allowed).\\

\midrule

\rowcolor{gray!20}\textbf{ReasonRank Output}\\
\midrule
\green{<think>}
Okay, let me try to figure out how to rank these passages based on the query about the longest increasing path in a matrix. The user is looking for the length of the longest path.

First, I need to go through each passage and see which ones are relevant. The query is specifically about dynamic programming approaches for this problem, possibly using memoization since it's a grid traversal with constraints.

Passage [3] talks about the longest increasing subsequence using dynamic programming. That's related but not exactly the same as the matrix problem. It's about sequences, not paths in a grid. So maybe not the top. Passage [6] is the length of the longest increasing subsequence again. Similar to [3], so not directly relevant. Passage [13] discusses maximal square area, which uses DP but not for paths. Not relevant. Passage [16] is about the longest palindromic subsequence. Again, subsequence, not grid paths. Not relevant. Passage [19] deals with unique paths with obstacles using DP. It's about grid paths but not increasing paths. Might be somewhat related but not directly. Passage [11] is shortest clear path in a binary matrix using BFS. Different problem but grid paths. Maybe somewhat related but not the same. Passage [15] involves collecting gold in a grid with DFS. It's about grid traversal but not increasing paths. Not directly relevant. Passage [20] is about longest increasing subsequence with difference constraints. Subsequence again, not grid. Not relevant. Passage [4] counts the number of longest increasing subsequences. Still subsequence, not grid paths. Passage [9] is longest common subsequence. Not relevant. Passage [10] is LCS again. Not applicable. Passage [14] involves min operations for robots on a 2xN grid. Not related to increasing paths. Passage [5] is about Excel cells, not relevant. Passage [18] is about work hours and palindromic intervals. Not related. Passage [2] is about strictly increasing paths in a grid, returning the count. Passage [1] discusses max points in queries starting from top-left, using DFS. It involves grid traversal and paths but not the specific increasing path problem. Might be somewhat related but not directly. Passage [17] is about LCS length again. Not relevant.

So the most relevant is passage [2], which directly addresses strictly increasing paths in a grid. Then passage [19] and [8] deal with grid paths but not increasing. Passage [3] and [6] are about subsequences, not paths. Passage [15] is about grid traversal but not increasing. 

Wait, passage [2] has code that seems to count strictly increasing paths, which is exactly the problem. The example in [2] matches the problem's examples. So [2] is top. Then maybe [19] and [8] for grid DP, but they're not directly solving the same problem. 

Passage [20] is about subsequences with difference constraints, not paths. So the ranking would have [2] first, then perhaps [19], [8], [7], etc., but the key is [2] is the most relevant. Then maybe [3] and [6] since they handle increasing sequences, but not grid paths. 

So the order should start with [2], then others that involve grid paths or DP for paths. But the query is specifically about the longest increasing path in a matrix. So [2] is the most relevant. Then maybe [19] and [8] for grid DP, but they're not the same. 

So the top passages are [2], followed by [19], [8], [7], [15], [1], etc. But need to check each again.

Looking at passage [20], it's about subsequences, not paths. Passage [13] is about squares, not paths. Passage [16] is LCS. So the most relevant is [2], then maybe [19] and [8] for grid DP. 

But the problem in the query is a graph problem. Passage [2] might use a similar approach. The code in [2] seems to count paths, which could be part of the solution. 

So the ranking would be:

[2] > [19] > [8] > [7] > [15] > [1] > [3] > [6] > [20] > [4] > [9] > [10] > [13] > [16] > [17] > [11] > [14] > [5] > [18] > [12]

But need to verify. The problem in the query is a classic DP with memoization on the grid. Passage [2] might be about counting all paths, which is a different problem but uses similar techniques. However, the code in [2] might not directly solve the longest path. 

Alternatively, passage [19] is about unique paths with obstacles, which uses DP but not for increasing paths. Passage [8] is min path sum, which uses DP but again not for increasing. 

So the most relevant is [2], then maybe [19], [8], [7], [15], [1], etc. But I'm not sure. The query's problem is a graph traversal with DP, so passages that deal with grid paths and DP would be more relevant. 

Passage [2] is about strictly increasing paths, which is exactly the problem. So [2] is definitely the top. Then perhaps [19] and [8] for grid DP. 

So the final ranking would start with [2], then others that involve grid paths or DP for paths. The rest are less relevant.
\green{</think>}

\purple{<answer>} 

[2] > [19] > [8] > [7] > [15] > [1] > [3] > [6] > [20] > [4] > [9] > [10] > [13] > [16] > [17] > [11] > [14] > [5] > [18] > [12]

\purple{</answer>}
\\
\bottomrule
\end{tabular}
\label{tab:LeetCode} 
\end{table*}

\begin{table*}[!t]
\centering
\caption{Case on AoPS. We can see that the model first analyzes the search query. Then, it looks at each passage and judges their relevance. Finally, it takes a closer look at the most relevant ones and makes a detailed comparison.}
\fontsize{9pt}{11pt}\selectfont
\begin{tabular}{p{0.98\linewidth}}
\midrule
\rowcolor{gray!20}\textbf{Query} \\
\midrule
Triangle $ABC$ is inscribed in circle $\omega$. Points $P$ and $Q$ are on side $\overline{AB}$ with $AP<AQ$. Rays $CP$ and $CQ$ meet $\omega$ again at $S$ and $T$ (other than $C$), respectively. If $AP=4,PQ=3,QB=6,BT=5,$ and $AS=7$, then $ST=\frac{m}{n}$, where $m$ and $n$ are relatively prime positive integers. Find $m+n$.\\
\midrule

\rowcolor{gray!20}\textbf{ReasonRank Output}\\
\midrule
\green{<think>}

Okay, let me try to figure out how to rank these passages for the given problem. The query is about a triangle inscribed in a circle with some points and intersections. The problem involves calculating ST given various lengths. 

First, I need to understand what the problem is asking. It's a geometry problem involving a circle, triangle ABC, points P and Q on AB, rays CP and CQ intersecting the circle again at S and T, and given specific segment lengths like AP=4, PQ=3, QB=6, BT=5, AS=7. The goal is to find ST as a fraction m/n and then m+n.

Now, I need to rank the 20 passages based on relevance. The most relevant ones would be those that deal with similar geometry problems, especially involving circles, triangles, intersecting lines, power of a point, chords, or Ceva's theorem, Menelaus, etc.

Looking at the passages:

[1] Talks about a circle, chord lengths, power of a point. Might be relevant since the problem involves chords and circle properties.

[2] Involves triangle inscribed in a circle, tangents, and intersections. Could be relevant as it's about triangle geometry with circle.

[3] Circles intersecting, common tangents, trapezoid area. Less relevant, seems more about trapezoids and tangents.

[4] Circumcenter and distance calculation. Maybe somewhat related but not directly.

[5] Incircle trisecting median. Not directly related.

[6] Tangent from origin to circle. Maybe not.

[7] Parabola and circle, tangent length. Not relevant.

[8] Parallelogram and circle tangent. Not relevant.

[9] Circle tangent to other circle. Not relevant.

[10] Perimeter with Law of Cosines. Maybe trigonometry, but not directly.

[11] Circumference calculation. Not relevant.

[12] Circumcircle and tangents. Could be relevant as it involves circumcircle and intersections.

[13] Point inside triangle with congruent angles. Not directly related.

[14] Concentric circles and tangents. Not relevant.

[15] Circle tangent to square. Not relevant.

[16] Hexagon inscribed in circle. Not relevant.

[17] Tangent and angle bisector. Maybe involves some circle theorems.

[18] Similar to [14], concentric circles. Not relevant.

[19] Right triangle legs. Basic Pythagorean, not relevant.

[20] Circle tangent to parabola. Not relevant.

So the most relevant ones are probably [1], [2], [12], [17], maybe [4], [20]. 

Passage [1] discusses chord lengths and power of a point, which might be useful here since the problem involves ST as a chord. 

Passage [2] is about triangle inscribed in a circle with tangents and intersections, which is very similar to the query's setup.

Passage [12] involves circumcircle and tangent points, which could relate to the problem's intersections at S and T.

Passage [17] deals with tangents and angle bisectors, which might involve some theorems used in the problem.

Passage [4] uses circumradius and power of a point, which could be applicable.

Passage [20] involves circle tangents and equations, but maybe less so.

So the ranking would start with [2], then [1], [12], [17], [4], followed by others. Need to check which directly address circle geometry with intersecting chords, power of a point, or Ceva's/Menelaus.
\green{</think>}

\purple{<answer>} 

[2] > [1] > [12] > [17] > [4] > [20] > [8] > [14] > [18] > [5] > [3] > [6] > [7] > [10] > [13] > [9] > [15] > [16] > [19] > [11] 

\purple{</answer>} \\
\bottomrule
\end{tabular}
\label{tab:AoPS} 
\end{table*}

\begin{table*}[!t]
\centering
\caption{Case study on TheoremQA-Theorems. The model first attempts to solve the problem, then checks the candidate passages.}
\fontsize{9pt}{11pt}\selectfont
\begin{tabular}{p{0.98\linewidth}}
\midrule
\rowcolor{gray!20}\textbf{Query} \\
\midrule
You are working with four possible machines, each of which is represented by a matrix that operates on an input vector $v$: (a) $(\Phi^T\Phi)^{-1}$. (b) $\Phi(\Phi^T\Phi)^{-1}$. (c) $\Phi(\Phi^T\Phi)^{-1}\Phi^T$. (d) $\Phi^T(\Phi^T\Phi)^{-1}\Phi^T$. Which of them takes any vector $v$ and projects it onto the space spanned by the columns of $\Phi$?
\\

\midrule

\rowcolor{gray!20}\textbf{ReasonRank Output}\\
\midrule
\green{<think>}
Okay, let me try to figure out how to rank these passages for the query. The user is asking which of the four matrices projects any vector $\mathbf{v}$ onto the space spanned by the columns of $\Phi$. The options are:

\begin{itemize}
    \item[(a)] $(\Phi^T \Phi)^{-1}$
    \item[(b)] $\Phi (\Phi^T \Phi)^{-1}$
    \item[(c)] $\Phi (\Phi^T \Phi)^{-1} \Phi^T$
    \item[(d)] $\Phi^T (\Phi^T \Phi)^{-1} \Phi^T$
\end{itemize}

First, I need to recall what an orthogonal projection matrix looks like. From linear algebra, the projection matrix onto the column space of $\Phi$ is usually given by $\Phi (\Phi^T \Phi)^{-1} \Phi^T$. So options (c) and (d) might be candidates. But let me check each one.

Option (a) is just the inverse of $\Phi^T \Phi$, which isn't a projection matrix. Projection matrices must be idempotent ($P^2 = P$) and symmetric. (a) is likely not idempotent.

Option (b) is $\Phi$ multiplied by that inverse. Multiplying $\Phi$ by that matrix might not yield a projection. Let me think: $\Phi (\Phi^T \Phi)^{-1}$ is the Moore-Penrose pseudoinverse of $\Phi$, but the projection is $\Phi$ times that times $\Phi^T$. So (c) is $\Phi \Phi^{\dagger} \Phi^T$, which is the projection.

Option (d) is $\Phi^T$ pseudoinverse times $\Phi^T$. Wait, no—that's not right. The correct form is $\Phi$ pseudoinverse times $\Phi^T$. So (c) is correct. Option (d), written as $\Phi^T (\Phi^T \Phi)^{-1} \Phi^T$, is not the same. Therefore, the correct answer is (c).

Now, looking at the passages, which ones discuss orthogonal projections onto the column space? Passage [2] talks about orthogonal projection onto the closed linear span using an orthonormal basis, which is exactly what's needed here. Passage [3] mentions orthogonal projections as linear transformations, which is relevant. Passage [8] discusses properties of projections, like P composed with P is P. Passage [17] defines projections in Hilbert spaces, which includes the kernel condition. Passage [20] talks about bounded projections, which might be related but less directly. Passage [6] explains linear transformations as matrix products, which is relevant for understanding how these matrices act. Passage [9] involves inner products and matrix factors, which could relate to the projection formula. Passage [14] is about matrix products as linear transformations, which is key here. 

So the most relevant passages would be those that explain orthogonal projections, especially using the formula $\Phi (\Phi^T \Phi)^{-1} \Phi^T$. 

Passage [2] gives the formula for orthogonal projection onto a subspace. Passage [3] confirms it's a linear transformation. Passage [8] discusses idempotence, a property of projections. Passage [17] defines projections. Passage [6] connects matrices to linear transformations. 

Others, like [9] and [14], might offer supporting information but are less directly relevant.

So the ranking would start with [2], then [3], [8], [17], [6], followed by others like [20], [9], [14], etc. Need to check which passages explicitly mention the projection formula or idempotence.
\green{</think>}

\purple{<answer>} 

[2] > [3] > [8] > [17] > [6] > [20] > [9] > [14] > [12] > [1] > [7] > [10] > [18] > [13] > [15] > [4] > [5] > [11] > [16] > [19]

\purple{</answer>}
\\
\bottomrule
\end{tabular}
\label{tab:theoremqa-t} 
\end{table*}

\end{document}